\documentclass{aa}  

\usepackage{graphicx}
\usepackage{hyperref}
\usepackage{txfonts}
\usepackage{orcidlink}
\usepackage{natbib}
\bibpunct{(}{)}{;}{a}{}{,} 
\usepackage{xcolor} 
\usepackage{booktabs} 
\usepackage{listings} 
\usepackage{csquotes} 
\usepackage[htt]{hyphenat} 

\definecolor{codegreen}{rgb}{0,0.6,0}
\definecolor{codegray}{rgb}{0.5,0.5,0.5}
\definecolor{codepurple}{rgb}{0.58,0,0.82}
\definecolor{backcolour}{rgb}{0.95,0.95,0.92}

\lstdefinestyle{mystyle}{
backgroundcolor=\color{backcolour},
commentstyle=\color{codegreen},
keywordstyle=\color{magenta},
numberstyle=\tiny\color{codegray},
stringstyle=\color{codepurple},
basicstyle=\ttfamily\footnotesize,
breakatwhitespace=false,
breaklines=true,
captionpos=b,
keepspaces=true,
numbers=left,
numbersep=5pt,
showspaces=false,
showstringspaces=false,
showtabs=false,
tabsize=2}
\usepackage{pifont}

\newcommand{\eg}{e.g.\@}
\newcommand{\cf}{cf.\@}

\newcommand{\km}{\mathrm{km}}

\newcommand{\mpc}{\mathrm{Mpc}}
\newcommand{\kpc}{\mathrm{kpc}}

\newcommand{\s}{\mathrm{s}}

\newcommand{\erg}{\mathrm{erg}}

\newcommand{\kms}{\km\,\s^{-1}}

\newcommand{\rsearch}{r_{\mathrm{search}}}
\newcommand{\mbh}{M_\bullet}
\newcommand{\rtd}{r_\mathrm{td}}
\newcommand{\mstar}{M_\star}
\newcommand{\rstar}{R_\star}
\newcommand{\msol}{M_\odot}
\newcommand{\rmb}{r_\mathrm{mb}}
\newcommand{\atilde}{\tilde{a}}
\newcommand{\lstar}{L_\star}
\newcommand{\yr}{\mathrm{yr}}
\newcommand{\fa}{\mathrm{FA}}
\newcommand{\gvol}{\Gamma_\mathrm{vol}}
\newcommand{\gabs}{\Gamma_\mathrm{abs}}
\newcommand{\gpc}{\mathrm{Gpc}}
\newcommand{\pyrpgal}{\yr^{-1}\,\mathrm{gal}^{-1}}

\begin{document} 

   \title{Rates of tidal disruption events from constrained cosmological simulations of the local Universe: population properties and implications for transient surveys}

   \titlerunning{Rates of tidal disruption events from constrained cosmological simulations of the local Universe}

   \author{
      Julian S. Sommer\inst{1}\,\orcidlink{0000-0002-1154-8317},
      Ildar Khabibullin\inst{2,1,3}\,\orcidlink{0000-0003-3701-5882},
      Klaus Dolag\inst{1,3},
      Luca Sala\inst{1,4}\,\orcidlink{0000-0002-5416-6394},
      Benjamin Seidel\inst{1}\,\orcidlink{0009-0000-3688-4379},
      Jenny G. Sorce\inst{5},
      Alice Damiano\inst{6}\,\orcidlink{0009-0000-5406-1333}
      }
\authorrunning{Sommer et al.}

   \institute{Universit\"ats-Sternwarte, Fakult\"at f\"ur  Physik, Ludwig-Maximilians Universität, Scheinerstr. 1, 81679 M\"unchen, Germany\\
   \email{julian.sommer@lmu.de}
         \and
         Rudolf Peierls Centre for Theoretical Physics, Department of Physics, University of Oxford, Clarendon Laboratory, Parks Rd, Oxford, OX1 3PU, United Kingdom
         \and
         Max-Planck-Institut für Astrophysik, Karl-Schwarzschild-Straße 1, 85741 Garching, Germany
         \and
         Excellence Cluster ORIGINS, Boltzmannstraße 2, 85748 Garching, Germany
         \and
         Univ. Lille, CNRS, Centrale Lille, UMR 9189 CRIStAL, 59000 Lille, France
         \and
         Dipartimento di Fisica dell’Università di Trieste, Sez. di Astronomia, via Tiepolo 11, 34131 Trieste, Italy
   }

   \date{\today}

 
  \abstract
   {With the new era of observational transient astronomy driven by the Legacy Survey of Space and Time (LSST) and recent advances in cosmological simulations, refined estimates for the rates of tidal disruption events (TDE) are of high relevance. Here, we provide TDE rate estimates based on the constrained cosmological Simulation of the LOcal Web (SLOW).}
   {Our goal is to evaluate the TDE rates within a fully 3D cosmological framework to test the limitations of traditional 2D analytical extrapolations. We aim to provide reliable TDE budgets extracted from the simulated zoom-in volumes of the digital counterparts of the Coma, Hercules, Shapley, Virgo, and Perseus supercluster environments, and the Fornax galaxy cluster.}
   {From the zoom-in boundary volumes of the six environments, reaching radial extents of $55-92\,\mpc$, we extracted black hole demographics (including spin) and their host galaxy properties to establish a filter scheme that strictly preserves dynamically stable \enquote{main-sequence} black holes. We further classified host galaxies as cuspy or cored based on the slope of their 3D stellar density profile measured within $1\,\kpc$ as a proxy for unresolved nuclear structure and applied the relativistic Kesden efficiency correction to the filtered sample.}
   {We find an average volumetric TDE rate of $\approx 600\,\gpc^{-3}\,\yr^{-1}$ across all six environments and find an average TDE rate per black hole of approximately $4.5\times 10^{-5}\,\yr^{-1}$. Although our absolute TDE rates match early literature estimates, the underlying spatial distribution fundamentally differs. Central core rates are heavily reduced by dynamical depletion and direct capture constraints, meaning the total TDE budget is overwhelmingly dominated by cuspy satellite galaxies in the extended cluster halos.}
   {TDE yields are driven by black hole demographics and spatial concentration rather than total cluster mass. Actively assembling superclusters, such as Hercules, systematically reduce per-black-hole TDE efficiencies due to merger-driven black hole mass growth. Low-mass environments, like Fornax, on the other hand, can be the most efficient per black hole due to a higher fraction of unmerged, low-mass black holes.}


   \maketitle

\section{Introduction}\label{sec:introduction}

Tidal disruption events (TDEs) are events that occur when stars encounter supermassive black holes (SMBHs) at a distance smaller than the tidal radius $r_\mathrm{td}$, where the differential gravitational forces of the black hole exceed the star's gravity \citep{Hills1975,1979SvAL....5...16L}. As was realized early on, accretion of matter of the disrupted star should lead to a bright flare of radiation lasting for months or years, depending on the exact configuration of the disruption \citep{1981A&A....95...39G,Rees1988,1989IAUS..136..543P,Evans1989}.

While supermassive black holes are ubiquitous in galactic centers, the vast majority lack persistent accretion and remain electromagnetically dark \citep{Kewley2006, Aird2012}. TDEs are therefore crucial transient probes that allow us to study the demographics of these otherwise quiescent black hole populations. By developing theoretical models to fit multi-wavelength light curves, we can infer fundamental parameters such as the black hole mass and spin directly from observations \citep{Angus2026, Ryu2020, Mockler2019, Mummery2024}. High-cadence photometric measurements from wide-field surveys are already uncovering TDE populations in the local universe \citep{Yao2023, Srivastav2026}. This is about to change with the advent of the Vera C. Rubin Observatory's Legacy Survey of Space and Time (LSST), which is expected to discover thousands of new TDEs each year \citep{Bricman2020, French2026}.

However, not all SMBHs can produce observable TDEs. Non-rotating black holes exceeding a critical mass limit (the Hills limit) fail to tidally disrupt stars, because the Schwarzschild radius $r_s$ grows linearly with the black hole's mass $M_\bullet$, whereas $r_\mathrm{td}\propto M_\bullet^{1/3}$. Stars in the vicinity of black holes above a black hole mass of $M\gtrsim 3\times 10^8\,\msol$ should result in direct captures.

Spinning SMBHs can push the Hills mass limit further up \citep[e.g.,][]{Ivanov2006}, because the marginally bound radius $\rmb$ (i.e., zero radial velocity and acceleration in Boyer-Lindquist coordinates \citep{Boyer1967}) can be reached at shorter distances for the same $M_\bullet$ if the spin is introduced. For example, for a black hole rotating with the maximum spin, the Hills mass limit is pushed to $7\times 10^8\,\msol$ \citep{Kesden2012}. To predict how efficiently black holes can produce observable flares, one has to infer the exact range where a star is marginally bound in its periapsis while staying within the tidal radius ($r_\mathrm{mb}<r<r_\mathrm{td}$). This directly translates into an efficiency parameter, called Kesden efficiency $\eta$ in what follows, which quantifies the exact fraction of the loss cone capable of producing TDEs relative to direct captures.

To statistically predict the rates at which black holes encounter close enough interactions with stars, one has to analyze the phase space of star clusters around black holes that result in either direct captures or TDEs \citep[see][for a recent review]{Stone2020}. 
The key parameter of the stellar orbit is its angular momentum, and it should be within a so-called loss-cone for a close enough encounter to occur \citep{Frank1976}. The influx of such stars, which is described by the loss-cone dynamics, highly depends on the assumed stellar density distribution function and velocity dispersion around the black hole \citep{Stone2020}. Theoretical predictions from two-body relaxation calculations typically yield rates $\gtrsim 10^{-4}\,\pyrpgal$ \citep{Magorrian1999, Wang2004}, but observed rates are generally an order of magnitude lower \citep{Stone2016}.

Since TDE rates strongly depend on the density profile of the surrounding star cluster, it is crucial to distinguish cuspy stellar density profiles from cored galaxies \citep{Stone2016, Stone2020}. Consequently, different baseline TDE rates for these two galactic environments were applied depending on the inner slope of the density profile $g$, where cusp-galaxies ($g\geq 2$) are more efficient in producing TDEs than core-galaxies ($g\leq 1$). This study builds upon these different baseline rates ($\Gamma_{g\leq 1}$ and $\Gamma_{g\geq 1}$), which are coupled to empirical data \citep{McConnell2013}. Recent studies \citep[e.g.][]{Hannah2024, Hannah2025} leverage this dichotomy by applying detailed loss-cone dynamics models to highly resolved density profiles of 91 galaxies with the \texttt{REPTiDE}-package. Besides confirming that stellar dynamics drive the loss-cone flux, they found that TDE rates are peaking in galaxies with masses of $10^{9.5}\,\msol$ and black holes with masses of $10^{6.5}\,\msol$.

Complementing these resolved, galaxy-by-galaxy studies, semi-analytical models provide the broader cosmological context across cosmic time. Combining the \textit{L-Galaxies} semi-analytical model with time-dependent loss-cone dynamics, \cite{Polkas2024} showed that nuclear star clusters dominate the TDE budget and that matching observed rates requires a high black-hole occupation fraction even in dwarf galaxies. \cite{Chowdhury2026} computed per-galaxy TDE rates from \textit{FIRE-2} cosmological zoom-in simulations across $z=1-10$, finding that rates peak at $\sim 4\times 10^{-4}\,\yr^{-1}$ near $z\sim 2.5$ and decline to $\sim 10^{-5}\,\yr^{-1}$ at $z=1$. Neither resolves how cluster assembly history and spatial concentration modulate TDE production in individual cluster environments.

There is a significant discrepancy between theoretical predictions of the TDE rate and observational measurements. In contrast to the rates of $\gtrsim 10^{-4}\,\yr^{-1}\,\mathrm{gal}^{-1}$ predicted for low-mass cuspy galaxies by idealized models of two-body relaxation models \citep{Magorrian1999, Wang2004}, global observed rates are found to be lower by approximately one order of magnitude. Early empirical models such as \cite{vanVelzen2014} and \cite{Holoien2016} find TDE rates of $(1.5-2.0)^{+2.7}_{-1.3}\times 10^{-5}\,\pyrpgal$ and $(2.2-17.0)\times 10^{-5}\,\pyrpgal$, respectively, in optical data. A more recent optical study by \cite{Yao2023}, based on data from the Zwicky Transient Facility (ZTF), proposes a TDE rate of $3.2\times 10^{-5}\,\pyrpgal$, where the TDE rate was inferred from a sample of 33 TDEs. This flux-limited, spectroscopically complete sample yields a volumetric TDE rate of $310^{+60}_{-100}\,\gpc^{-3}\,\yr^{-1}$ for TDEs with peak blackbody luminosities $L_\mathrm{bb}>10^{43}\,\erg\,\s^{-1}$. Furthermore, X-ray all-sky surveys have been a foundational tool for finding TDEs \citep[e.g.][]{2015JHEAp...7..148K}, where, for example, candidates found by cross-correlating ROSAT and XMM-Newton observations imply a TDE rate of $\sim 3\times 10^{-5}\,\pyrpgal$ \citep{Khabibullin2014b}. The potential of the SRG/eROSITA mission to detect TDEs was theoretically predicted well in advance \citep{Khabibullin2014a}. TDE rates found by the eROSITA consortium covering the eastern \citep{Sazonov2021} and western galactic hemisphere \citep{Grotova2025} are in strong agreement, reporting $(1.1\pm 0.5)\times 10^{-5}\,\pyrpgal$ and $\approx 1.2\times 10^{-5}\,\pyrpgal$, respectively. However, the slight discrepancy between optical and X-ray data could point to obscuration at different viewing angles or to different emission processes peaking at different wavelengths. Dust obscuration can bias the TDE rate found in optical data. \cite{Masterson2024} studied a selection of TDE candidates that had no optical counterpart using NEOWISE mid-IR data and placed a lower limit on the TDE rate estimate at $(2.0\pm 0.3)\times 10^{-5}\,\pyrpgal$. Extending this approach to the high luminosity regime ($L_\mathrm{peak,\,W2} \gtrsim 3\times 10^{43}\,\mathrm{erg}\,\s^{-1}$, \cite{Nair2026} identified a luminosity function break at $\log(L_\mathrm{peak,\,W2}/\mathrm{erg}\,\s^{-1})\simeq 43.4$, where high-luminosity events are suppressed by more than 200 times relative to low-luminosity extrapolations. They attribute this turnover to the tidal radius falling inside the event horizon for non-spinning black holes with $\mbh\approx 10^8\,\msol$, providing direct observational evidence for the black hole-mass suppression.

The discrepancies mentioned above highlight the limitations of simple theoretical models. TDE rates depend strongly on galaxy evolution, shaped by the complex interplay between cosmological structure formation, star formation, and SMBH feedback processes. The use of highly resolved cosmological simulations is therefore a promising way to improve on these models. In this study, we analyze simulated zoom-in regions of five local supercluster environments (Coma, Hercules, Shapley, Virgo, Perseus) and one isolated galaxy cluster (Fornax), where galaxies at $z=0$ are the direct result of this complex structure-formation process. This enables us to study TDE rates in cosmologically evolved galaxies that represent realistic black hole demographics, while still resolving their neighboring stellar environments sufficiently to fully analyze loss-cone dynamics. This helps us to further understand the TDE rates and spatial distribution of TDEs in cluster environments, where, for example, dynamical core depletion becomes a dominant factor.

\section{Methodology}\label{sec:methodology}

We outline the simulation framework and zoom-in setup (Section \ref{subsec:simulation}), define the volume boundaries and black hole extraction procedure (Section \ref{subsec:bhs}), and describe how we characterize host environments through stellar density slopes, velocity dispersion, and fractional anisotropy, resulting in the four main-sequence filters that isolate dynamically stable black holes (Section \ref{subsec:stellar_density}). We then derive the relativistic Kesden efficiency correction (Section \ref{subsec:rel_constraints}) before calculating volumetric TDE rates using the SM16 framework \citep{Stone2016} (Section \ref{subsec:vol_tde_rates}).

\subsection{The cosmological simulations}\label{subsec:simulation}

The basis of our work is set by the state-of-the-art constrained cosmological simulation of the Local Universe called "Simulation of the LOcal Web" \citep[SLOW][]{Dolag2023}.

\subsubsection{The SLOW parent simulation box}

The initial conditions for SLOW are based on the approach using galaxy peculiar velocities, which is described in detail by \citet{sorce18}. A summary of the most important steps in building these initial conditions, as well as a detailed description of the cross-identified structures and clusters, is summarized in \citet{Dolag2023}, \citet{hernandezmartinez+24}, and \citet{Seidel2025}. The observational dataset underlying the initial conditions is the Cosmicflows-2 catalog \citep[CF2,][]{Tully2013}. The resulting initial conditions are based on the CLONE \citep[Constrained LOcal \& Nesting Environment,][]{sorce+21, sorce+24} simulations (in this case, realization number 8). SLOW covers a volume of $(500\,h^{-1}\,\mpc)^3$ and assumes cosmology based on \cite{Planck2014}, with a Hubble constant $H_0 = 67.77\kms\,\mpc^{-1}$, baryon fraction $\Omega_B = 0.0480217$, total matter density $\Omega_M=0.307115$, cosmological constant $\Omega_\Lambda = 0.692885$, normalization of the power spectrum $\sigma_8 = 0.829$ and a slope of the primordial fluctuation spectra $n = 0.961$.

\subsubsection{The high resolution zoom in regions}

The ``LOcal WEb Re-simulations with Dynamical friction and Extended blaCK hole Spin model'' (LOWER DECKS) are re-simulations of a large number of specially selected regions at a particularly high resolution \citep{Seidel2026}, capable of well resolving galaxies down to stellar masses of $M_*=10^{9}M_\odot$, with particle masses of $m_\mathrm{dm}=5.7\times10^{7} M_{\odot}$, $m_\mathrm{gas} = 1.0\times10^{7} M_{\odot}$ and $m_\star = 2.5\times10^{6} M_{\odot}$, for the dark matter, gas and stellar particles, respectively. The comoving gravitational softening lengths are set to $\epsilon_\mathrm{dm}=2.25\,h^{-1}\kpc$, $\epsilon_\mathrm{gas}=0.75\,h^{-1}\kpc$, and $\epsilon_\star=\epsilon_\bullet=0.35\,h^{-1}\kpc$. To prevent unphysical growth at late times, these are capped at maximum physical softening limits of $\epsilon_\mathrm{dm, phys}=\epsilon_\mathrm{gas, phys}=0.75\,h^{-1}\kpc$ and $\epsilon_\star=\epsilon_\bullet=0.35\,h^{-1}\kpc$. Here, we used six zoom-in regions: five supercluster environments (Coma, Hercules, Shapley, Virgo, Perseus) and one galaxy cluster environment (Fornax). In this set $M_{500}$ spans from $3.5\times 10^{13}$ to $1.7 \times 10^{15}\,\msol$, covering two orders of magnitude from group-scale to massive supercluster environments. Each zoom-in region is much larger than the virialized volume of each anchor cluster today, ranging over $(63^3-93^3)\,\mpc^3$. 

\subsubsection{The galaxy formation model}

These simulations were conducted with Open\textsc{Gadget}3 \citep{Groth2023
} and are based mostly on the {\it Magneticum} \citep{2016MNRAS.463.1797D,Dolag2025} galaxy formation model. This features a sub-resolution description of star-formation \citep{SpringelHernquist2003}, metal depending cooling \citep{Wiersma2009,Tornatore2010}, stellar evolution \citep{Tornatore2004,Tornatore2007}, improved smoothed particle hydrodynamics \citep{Dehnen2012,Beck2016}, a black hole subgrid description \citep{SpringelDiMatteo2006,Fabjan2010,Hirschmann2014} and a treatment of isotropic, thermal conduction \citep{2014arXiv1412.6533A}.

Importantly, the underlying black hole model follows the growth and evolution of the spin of the individual black holes, applying a detailed description for how the angular momentum is transported from the accreted gas to the black holes and how the spin evolves during mergers of two black holes \citep[see][for details]{Sala2024}. In addition, the treatment of the black hole sink particles is improved by an updated treatment of the unresolved dynamical friction \citep[see][for details]{Damiano2024}. To identify galaxies in the simulations, we are using the \textsc{SubFind} algorithm \citep{Springel2001, Dolag2009}.

Subhalos identified by \textsc{SubFind} exhibit realistic macroscopic and internal properties that match observational scaling relations across cosmic time, including mass-size evolutions, metallicities, and quenching mechanisms \citep{Kudritzki2021, Kimmig2025, Dolag2025}. Additionally, the {\it Magneticum} model reproduces the observed morphological dichotomy between spheroidal and disk galaxies, exhibiting accurate angular momentum distributions \citep{Teklu2015, Teklu2017}. In this model, the internal stellar kinematics align very well with observations, specifically recovering the distinct structural differences between fast and slow rotators \citep{Schulze2018}. Overall, the simulation accurately captures complex environmental processes in cluster outskirts, such as the formation of tidal streams from disrupted satellites \citep{Stoiber2025} and the stripping of dwarf galaxies \citep{Ivleva2024}.

\subsection{Simulation volume and black hole extraction}\label{subsec:bhs}

The high-resolution zoom-in volumes are defined by a set of boundary particles \citep{Seidel2026}. To determine this boundary, the collapsed halo is identified in the far future of the simulation (corresponding to the scale factor $a=1000$), and its Friends-of-Friends (FOF) group is extracted. The physical turnaround radius is then defined as the boundary where the inward radial velocity transitions to zero. The particles defining this turnaround surface are traced back to the initial conditions to dictate the exact extent of the high-resolution region.

To construct a closed surface around the simulated zoom-in region, we generated a HEALPIX map \citep{Gorski2005} from these boundary particles in the ring ordering scheme. The radial distance of each pixel was defined by the position of the corresponding boundary particle, minus an inward buffer zone of $r_\mathrm{buffer}=1\,\mpc$ to mitigate edge artifacts. In rare cases where a pixel contained no boundary particle, we interpolated the distance using the mean of its eight neighboring pixels. All simulated black holes located within this bounded volume were extracted for analysis. Furthermore, because these resulting zoom-in regions have highly asymmetric boundaries, we defined a maximum normalization radius, $R_\mathrm{safe}$, for each individual environment to maintain consistent spatial comparisons. $R_\mathrm{safe}$ is defined by the radius of the largest sphere that can be inscribed within the buffered boundary surface. Establishing this inscribed spherical volume ensures that radial profiles can be compared between different zoom-in regions.

The volume of each zoom-in boundary region ($V_\mathrm{z}$) is computed by integrating over the HEALPix decomposition of the boundary surface. For each pixel $i$, the solid angle $\omega_\mathrm{pix}= 4\pi/N_\mathrm{pix}$ is multiplied by the radial extent ($r_i-r_\mathrm{buffer}$), yielding the differential volume element $\mathrm{d}V_i=(1/3)\omega_\mathrm{pix}(r_i-r_\mathrm{buffer})^3$. The total volume is given by the sum over all pixels.

For each selected black hole, we extracted its mass $\mbh$ and dimensionless spin parameter $\atilde$ from the simulation data.

\subsection{Determination of nuclear density profiles}\label{subsec:stellar_density}

Due to computational and observational limitations in resolving nuclear star clusters (NSCs) directly, previous studies estimating tidal disruption rates \citep[e.g.,][]{Kesden2012} often rely on idealized approximations. A common approach is to assume an isotropic distribution function within a singular isothermal sphere (SIS) density profile \citep{Wang2004} to model the rate at which stellar diffusion fills the loss cone. This method inherently couples the disruption rate to empirical black hole mass-velocity dispersion ($M_\bullet-\sigma$) scaling relations \citep[e.g.,][]{Schulze2011}. However, real galactic environments exhibit significant structural diversity that deviates from this simplified assumption. Therefore, to accurately model the true loss-cone refilling rate, it is crucial to account for the specific environmental properties of each individual host galaxy.

To capture this structural diversity, we characterized the local environment by measuring the 3D stellar density slope ($g$, where $\rho_\star(r)\propto r^{-g}$) around each black hole to distinguish cusp from core galaxies. Because the inner parsecs of the NSC are not resolved, we measured $g$ over a radial envelope extending to $1\,\kpc$. We justified this approach based on the paradigm that a galaxy's macroscopic structure reflects its global assembly history, which fundamentally dictates its subgrid nuclear profile \citep{Faber1997, Lauer2007}. Dissipative and gas-rich processes are responsible for highly concentrated bulges with steep macroscopic density gradients that continue into central cusps \citep[e.g.,][]{Faber1997, Kormendy2009}. Gas-poor major mergers, on the other hand, produce more diffuse macroscopic envelopes with flattened central cores by dynamically scouring stars from the nucleus through binary black hole interactions \citep{Begelman1980, Milosavljevic2001}. Therefore, a density slope measured from within larger radii than the extent of the NSC serves as a proper proxy for identifying the underlying morphological family and predicting the unresolved central structure. Figure \ref{fig:slope_rcap_map} shows a parameter sweep of the stellar density slopes as a function of $\rsearch$ around each black hole. For $\rsearch\lesssim 0.5\,\kpc$, artificial gravitational softening dominates because $\rsearch$ approaches the maximum physical stellar softening limit ($\epsilon_{\star,\mathrm{phys}}\approx 0.52\,\kpc$), resulting in the artificial suppression of the core-cusp dichotomy. At $\rsearch\approx 1\,\kpc$, the two distinct populations stabilize, validating our choice of setting $\rsearch = 1\,\kpc$. In addition, the stability map clearly shows that if $\rsearch$ becomes too large, the core-cusp populations become indistinguishable because the outer galactic disk and the extended diffuse stellar halo contaminate the unique structural signature of the central bulge.

To map the local stellar environments, we constructed a $k$-d tree using the \texttt{NearestNeighbors.jl} package \citep{Carlsson2022} to efficiently query the nearest stellar particles surrounding each black hole in the zoom-in volume. A stellar particle was considered a neighbor if its relative distance to the central black hole fell within the search radius $\rsearch = 1\,\kpc$.

Since surrounding stars are a required criterion for TDE rates, black holes were excluded from the TDE rate calculation if $N_\mathrm{neigh}<10$ were found within $\rsearch$. If $N_\mathrm{neigh}=0$, no stars could physically feed the black hole to produce a TDE, and if $0<N_\mathrm{neigh}<10$, not enough stellar particles would be available to fit a stellar density profile.

To determine the stellar density slope $g$ for each black hole environment, we calculated the density $\rho_\star$ in logarithmically spaced radial shell-bins extending out to $\rsearch$ and fit a linear regression to the $\log_{10}(\rho_\star)$ versus $\log_{10}(r)$ distribution, extracting $g$ as the negative slope to the fit. 

\begin{figure}
    \centering
    \includegraphics[width=\linewidth]{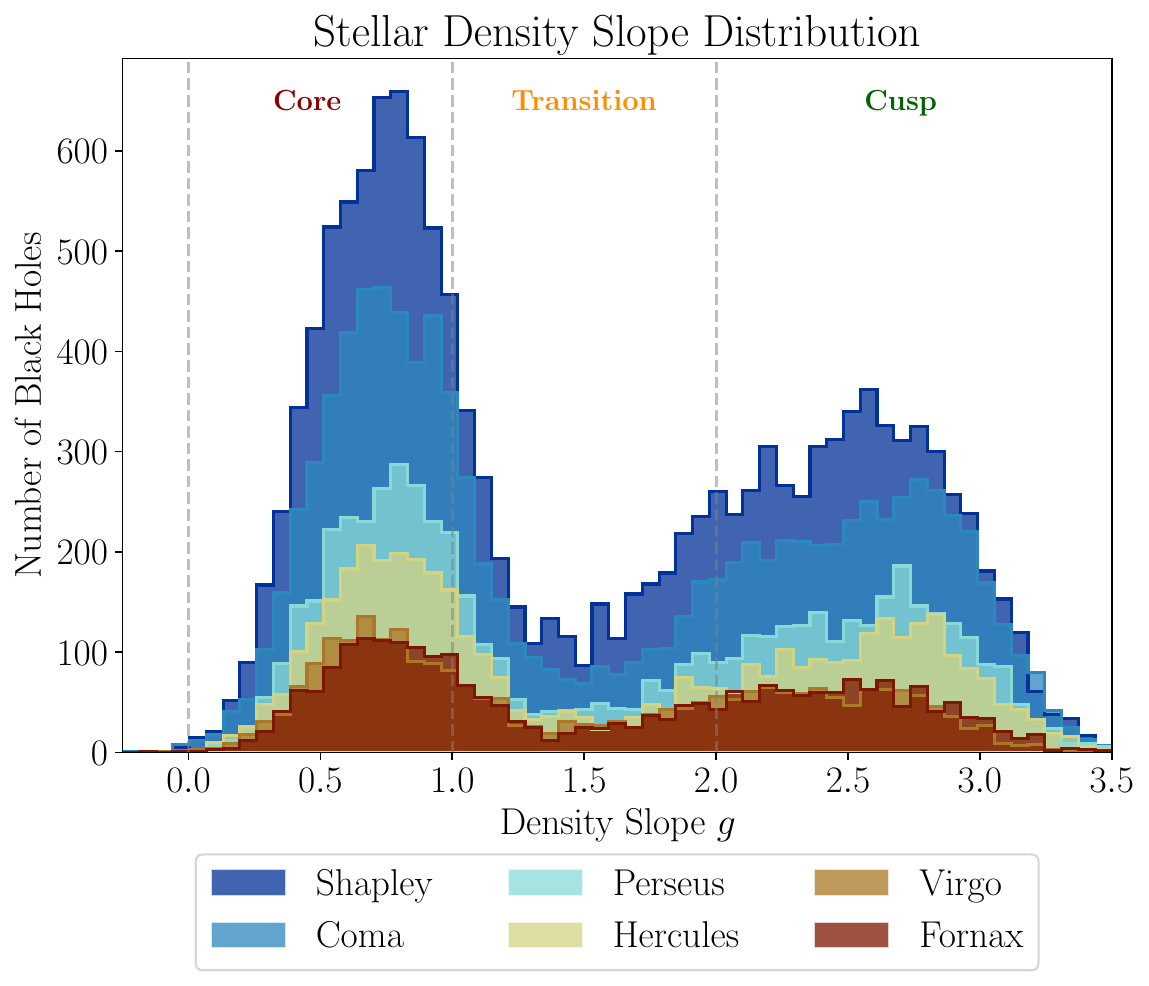}
    \caption{Distribution of stellar density slope $g$ for each valid black hole in the six simulated structures (Shapley, Coma, Perseus, Hercules, Virgo, and Fornax). A distinct bimodal distribution, separating core ($g \leq 1$) and cusp ($g \geq 2$) environments, is clearly observed across all environments.}
    \label{fig:gamma_distribution}
\end{figure}

Figure \ref{fig:gamma_distribution} shows the measured $g$-distribution around each valid black hole for the combined zoom-in volumes of the five supercluster regions and the Fornax cluster. The resulting distribution is strongly bimodal, with distinct populations peaking in the core ($g \leq 1$) and cusp ($g \geq 2$) regimes. This structural dichotomy is a well-established observational feature of galactic nuclei \citep{Faber1997, Lauer2007, Kormendy2009}, underscoring the physical importance of separating the loss-cone dynamics and applying distinct theoretical TDE rate equations for the core and cusp environments.

Massive galaxies often exhibit a break radius where profiles flatten inward and decline outward out to kiloparsec scales \citep[e.g.,][]{Faber1997}. Our use of logarithmic radial binning ensures that the kinematically flatter inner regions carry equal statistical weight. This prevents the steeper envelope outside the break radius from dominating the regression, allowing the globally fitted $g$-slope to robustly separate diffuse core systems from highly concentrated cusps.

To investigate the simulated host environments with observational scaling relations, we also evaluated the stellar velocity dispersion ($\sigma$). We computed the 3D root-mean-square velocity of all neighboring stellar particles relative to the central black hole's rest frame, then divided by $\sqrt{3}$ to obtain the equivalent 1D isotropic velocity dispersion. To establish a simple scalar metric to distinguish isotropic from anisotropic stellar environments, we use the fractional anisotropy (FA). To calculate FA, the 3D velocities of all stellar particles within $\rsearch$ were extracted, and the $3\times 3$ covariance matrix, representing the 3D velocity dispersion tensor, was computed. The three eigenvalues ($\lambda_1$, $\lambda_2$, $\lambda_3$) of this covariance matrix correspond to the velocity variances along the principal axes of the velocity ellipsoid. If $\lambda_1\approx \lambda_2\approx \lambda_3$, the velocity ellipsoid has the shape of a sphere, resembling an isotropic core. If one of the three eigenvalues is significantly smaller, the ellipsoid flattens. To measure how far the shape of the velocity ellipsoids differs from a perfect sphere, we use:
\begin{equation}
    \fa = \sqrt{\frac{1}{2}} \frac{\sqrt{(\lambda_1 - \lambda_2)^2 + (\lambda_2 - \lambda_3)^2 + (\lambda_3 - \lambda_1)^2}}{\sqrt{\lambda_1^2 + \lambda_2^2 + \lambda_3^2}},
\end{equation}
producing a scalar metric that scales between 0 and 1. If $\fa=0$, the eigenvalues are equal and the local environment is an isotropic, pressure-supported sphere. If $\fa=1$, one eigenvalue dominates the others and the local environment is highly anisotropic.

The loss-cone formalism and the empirical TDE rate prescriptions of SM16 are derived for black holes embedded in relaxed, pressure-supported, and approximately isotropic stellar environments. Rotation-dominated or highly anisotropic systems alter the phase-space filling of the loss cone and violate the conditions under which the SM16 prescriptions are calibrated. Furthermore, deriving the 1D velocity dispersion, a key input to the SM16 prescriptions, is physically meaningful only when the local velocity ellipsoid is approximately spherical. Applying the SM16 prescriptions to environments dominated by anisotropic velocity structures violates its foundational assumptions and introduces systematic biases \citep[see][for a discussion of anisotropic biases]{Stone2016}.

To isolate dynamically stable black hole environments sustaining isotropic loss-cone dynamics, we apply four main-sequence filters that exclude unbound, rotation-dominated, and tidally stripped systems:

\begin{enumerate}
    \item \textit{Offset cut:} Requires black holes to be located within a maximum distance of $1.5\,\kpc$ ($\approx 4\epsilon_\bullet$) from the center of their closest subhalo.
    \item \textit{Velocity cut:} The relative velocity between the black hole and the subhalo must be lower than the local stellar velocity dispersion within $\rsearch$, ensuring the black hole's motion being part of the local pressure-supported population.
    \item \textit{FA cut:} Restricts the stellar environments around black holes to those exhibiting a fractional anisotropy $\fa<0.4$ in their 3D velocity dispersion tensor.
    \item \textit{DM ratio cut:} To successfully identify and remove bare, tidally stripped stellar nuclei from intact galaxies, a dark matter to stellar mass ratio $>1$ is required from the subhalo.
\end{enumerate}

We have to acknowledge that the choice of the exact offset distance in filter 1 is motivated by the assumption that most SMBHs are located in the centers of galaxies. However, observations show that non-negligible exceptions of highly offset TDEs exist \citep[e.g., TDE 2025abcr with an offset of $9.08\pm 0.02\,\kpc$ from the host's nucleus,][]{Patra2026}. While such offset events exist, they represent rare configurations that do not reflect the sustained, dynamically stable environments required for reliable TDE rate estimation via the SM16 framework. Expanding the filter to such distances would contaminate the sample with black holes that do not contribute to the main sequence.

In dense cluster environments, tidal forces, mergers, and ram pressure can significantly perturb galaxies and distort their extended dark matter halos. Since the \textsc{subfind} algorithm \citep{Springel2001, Dolag2009} defines the global center of a subhalo as the position of the particles with minimum gravitational potential in conjunction with black holes in the SLOW simulation acting as free, collisionless sink particles \citep{Sala2024}, they can become offset from the global potential minimum during violent mergers. Beyond merger-driven offsets, low-mass black holes near their seeding mass are subject to poorly resolved dynamical friction. Close encounters with more massive particles heat their orbits, scattering them from their host centers even in quiescent environments. This numerical heating persists even with the sub-resolution dynamical friction correction of \cite{Damiano2024}, which, while significantly reducing spurious BH displacements compared to repositioning schemes, cannot fully recover analytical sinking timescales when $\mbh\lesssim M_\mathrm{particle}$. As shown by \citep{Damiano2025}, the timescale for low-mass BHs to return to the galactic center remains delayed, particularly in multi-component stellar-bulge systems where numerical heating is amplified. A single offset filter is therefore insufficient to confirm a valid TDE environment, and instead, considering the local kinematic stability of stars surrounding the black hole (filter 2 \& 3) provides a much more robust indicator that black holes truly reside within an intact, dynamically stable stellar core that is capable of sustaining the relevant loss-cone dynamics for TDEs.

\begin{figure*}
    \centering
    \sidecaption
    \includegraphics[width=0.66\linewidth,trim=0cm 0.5cm 0cm 0cm]{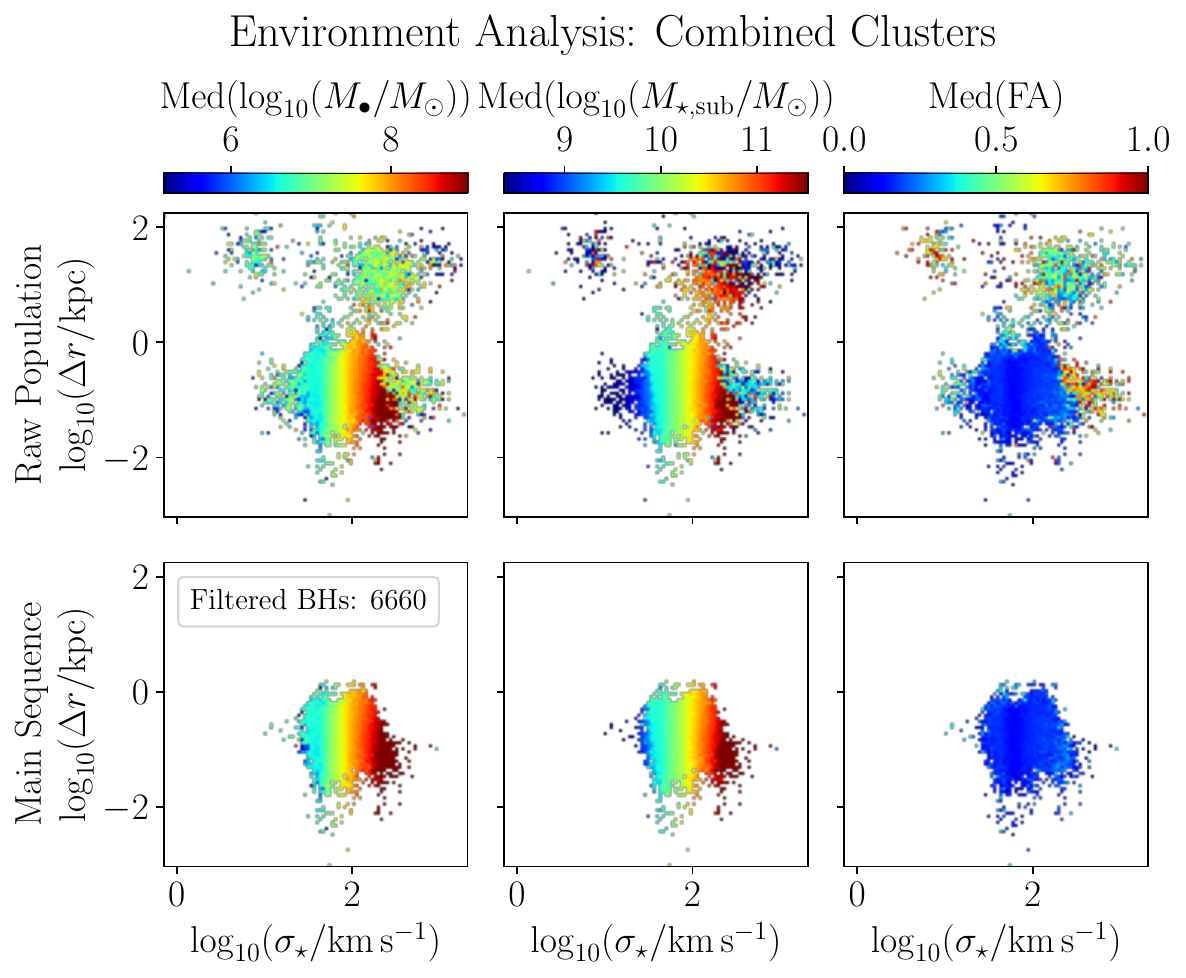}
    \caption{Population of black holes before (top row) and after the main-sequence-filters were applied (bottom row). Each column is color-coded by the median black hole mass ($\mbh$), stellar subhalo mass ($M_{\star,\mathrm{sub}}$) and fractional anisotropy (FA). The population in $\sigma_\star$-$\Delta r$ parameter space shows the velocity dispersion within the inner $1\,\kpc$ versus the offset of the black hole to its closest subhalo.}
    \label{fig:main_sequence_environment}
\end{figure*}

Figure \ref{fig:main_sequence_environment} shows the black hole population of all six environments in our sample, visualized in the relation between the projected velocity dispersion and the black hole offset to its closest subhalo. The three columns of subfigures are color-coded by the median black hole mass ($\mbh$), stellar mass of the subhalo ($\mstar$), and the fractional anisotropy (FA), respectively. The top row shows the raw population of black holes, while the bottom row depicts the population after all 4 filters were applied to the dataset. This figure shows that, after applying these filters, black holes survive in regions where the long-established $\mbh$-$\mstar$ and $\mbh$-$\sigma$ relationships hold \citep{Ferrarese2000, Gebhardt2000, Kormendy2013, McConnell2013, Reines2015}. Black holes that fulfill this relationship show almost entirely isotropic velocity dispersion in the central $1\,\kpc$.

\subsection{Relativistic constraints on tidal disruption efficiency}\label{subsec:rel_constraints}

In the Newtonian regime, a star with mass $M_\star$ and radius $R_\star$ in the vicinity of a static black hole is tidally disrupted at a distance $r<\rtd$, where the star's self-gravity equals the differential acceleration in the tidal field of the black hole $\rtd = \rstar (\mbh/\mstar)^{1/3}$.

In contrast to the tidal radius, the Schwarzschild radius $r_\mathrm{s}=2G\mbh/c^2$ for a non-spinning black hole scales linearly with $\mbh$, whereas $\rtd\propto \mbh^{1/3}$. By equating $\rtd$ and $r_\mathrm{s}$, one can find the maximum possible black hole mass capable of tidally disrupting a solar-like star, yielding $M_{\bullet,\,\mathrm{max}}\approx 1.1\times 10^8\,\msol$ \citep{Kesden2012}.

It is important to note that this approximation breaks down at the event horizon and when considering spinning Kerr black holes. Starting from the Kerr metric and considering prograde equatorial geodesics, the radius at which a star is on a marginally bound orbit, $\rmb$, can be found when the radial velocity and acceleration in Boyer-Lindquist coordinates \citep{Boyer1967} are zero. Following \cite{Bardeen1972, Kesden2012}, one finds the zone of no return:
\begin{equation}
    \rmb = \mbh \left( 1+\sqrt{1-\tilde{a}}\right)^2,
\end{equation}
with the dimensionless spin parameter $\atilde = a/\mbh$. Furthermore, \cite{Kesden2012} found that the general relativistic tidal disruption radius for Kerr black holes is given by the expression:
\begin{equation}\label{eq:r_td_relativistic}
    \rtd = \left(\frac{|\beta_{-}|}{\mbh/r^3}\frac{\mbh}{\mstar}\right)^{1/3}\rstar,
\end{equation}
where $\beta_{-}$ denotes the negative eigenvalue from the tidal tensor:
\begin{equation}
    \beta_{-}=-2\left(1+\frac{3K}{2r^2}\right)\frac{\mbh}{r^3},
\end{equation}
with $K=(L-aE)+Q$, where the Carter constant $Q$ is zero for equatorial geodesics and the specific energy $E$ can be set to unity for units where $c=1$. By equating $\rmb$ with the relativistic $\rtd$, one finds the maximum possible black hole mass capable of tidally disrupting solar-like stars:
\begin{equation}\label{eq:kerr_mmax}
    M_{\bullet,\,\mathrm{max}} = \sqrt{\frac{2\left(1+\frac{3\left(\tilde{L}_\mathrm{capt}-\atilde\right)^2}{2\tilde{r}_\mathrm{mb}^2}\right)}{\tilde{r}_\mathrm{mb}^3}\frac{\rstar^3}{\mstar}},
\end{equation}
with the dimensionless radius $\tilde{r}_\mathrm{mb}=\rmb/\mbh$ and the specific angular momentum along the $z$-axis $\tilde{L}_\mathrm{capt}=2\left(1+\sqrt{1-\atilde}\right)$, which is found by solving the double root while deriving $\rmb$.

In contrast to the Newtonian limit for non-rotating black holes, Equation \ref{eq:kerr_mmax} includes $\atilde$ as a free parameter. This explains why, for rotating black holes, the relativistic $M_{\bullet,\,\mathrm{max}}$ is pushed well beyond the standard Newtonian Hills mass limit.

However, to model how efficiently a black hole produces an observable flare, knowing only $\rmb$ and $\rtd$ is insufficient. For a successful tidal disruption event, a star must reach periapsis within the narrow radial window $\rmb<r<\rtd$. This condition strictly depends on the star's specific angular momentum, $\lstar$. If $\lstar<L_\mathrm{capt}$, the star will plunge directly into the event horizon without completing a disrupted orbit. Therefore, it is crucial to find the range $L_\mathrm{capt}<\lstar<L_\mathrm{td}$, which defines the specific wedge of angular momentum phase space (the loss cone) that allows a star to reach periapsis within $\rtd$. Following \cite{Kesden2012}, one can introduce an efficiency factor $\eta$ that filters the available loss cone, quantifying the true fraction of stars that result in an observable TDE rather than a direct capture.

Unlike in the Newtonian regime, the relativistic tidal disruption radius depends on the specific angular momentum. To satisfy the condition that the radial velocity at the periapsis must be zero, the solution for $L_\mathrm{td}$ depends on the radial distance. This establishes a circular dependency of $\rtd$ and $L_\mathrm{td}$. It is therefore not possible to analytically determine the maximum angular momentum at which tidal disruption remains possible, and an iterative approach must be used. We initialized our first guess with $\rtd$ from the Newtonian regime and calculated the specific angular momentum $L_i$ required for a marginally disrupting orbit. $L_i$ was then used to calculate the relativistic correction factor $|\beta_{-}|/(\mbh/r_i^3)$ to find $r_{\mathrm{td},i}$ until the fractional change between consecutive steps falls below $10^{-5}$. Once converged, $r_{\mathrm{td,\,final}}$ was inserted into the discriminant to find the final specific angular momentum $L_\mathrm{td}$. If $r_{\mathrm{td,\,final}}$ falls within the marginally bound capture radius ($\rmb$), the star is swallowed entirely, yielding an efficiency of zero. With $L_\mathrm{td}$, the fraction of the loss cone that results in an observable flare, while avoiding the direct capture region $L_\mathrm{capt}$, was found via $\eta = 1-(L_\mathrm{capt}/L_\mathrm{td})^2$.

The efficiency $\eta$ was applied as a multiplicative weight to the baseline TDE rate for every black hole, acting as a filter that naturally cuts off the rate for high-mass, low-spin systems. Figure \ref{fig:kesden_efficiency_map} shows the analytical mass limit, inferred from Equation \ref{eq:kerr_mmax}, embedded in the pixel grid of $\eta$ for different mass-spin combinations of the black hole.

\subsection{Calculation of volumetric TDE rates}\label{subsec:vol_tde_rates}

SM16 empirically found that the trend of the TDE rate depends on whether the black hole is located in a core or cusp environment:
\begin{align}\label{eq:stone_baserate_1}
    \Gamma_{g \leq 1} &= 1.2\times 10^{-5}\left(\frac{\mbh}{10^8\,\msol}\right)^{-0.247}\,\yr^{-1}\,\mathrm{gal}^{-1}\\
    \Gamma_{g \geq 2} &= 6.5\times 10^{-5}\left(\frac{\mbh}{10^8\,\msol}\right)^{-0.223}\,\yr^{-1}\,\mathrm{gal}^{-1}. \label{eq:stone_baserate_2}
\end{align}

We computed TDE rates using the SM16 framework, which assumes loss-cone refilling is driven by 2-body relaxation. While the simulation includes gas hydrodynamics, we did not account for gas-induced dynamical friction on stars in the rate calculation, consistent with the standard loss-cone formalism \citep[see][for a review]{Stone2020}.

To find the total volumetric TDE rate ($\Gamma_\mathrm{vol}$) in each simulated zoom-in volume, a sum over all $N_\bullet$ valid black holes was performed, where the individual base rates ($\Gamma_i$) were multiplied by their respective relativistic efficiency ($\eta_i$). The volumetric rate results from dividing this sum by the total volume of the high-resolution zoom-in region ($V_\mathrm{z}$):
\begin{equation}\label{eq:final_gamma_vol}
    \Gamma_\mathrm{vol} = \frac{1}{V_\mathrm{z}}\sum_{i=1}^{N_\bullet}\Gamma_i \eta_i.
\end{equation}

It is worth noting that the FA cut excludes highly rotating disks of stars within $\rsearch$, acting as a validity filter. For small galaxies where the resolution is too low to isolate the isotropic cores from their rotation-dominated disk on sub-kiloparsec scales, applying the SM16 framework would be mathematically invalid.

As a final remark, it is important to address a methodological nuance regarding the integration of these two frameworks in Equation \ref{eq:final_gamma_vol}. While SM16 incorporates TDE rates that are integrated over a full Kroupa stellar mass function, we calculated the relativistic efficiency $\eta$ assuming a fiducial solar-type star ($\mstar=1\,\msol$, $\rstar=1\,R_\odot$). Since $\rtd\propto \mstar^{-1/3}\rstar$, the theoretical zero-age main sequence mass-radius relation for the lower main sequence ($\rstar\propto\mstar^{0.79}$; \eg, \citep{Kippenhahn2013}) results in a weak mass dependence of $\rtd\propto\mstar^{0.46}$. Therefore, evaluating the event-horizon-capture geometry for a solar-mass star provides a robust limit on the survival of the loss-cone flux.

\section{Results}\label{sec:results}

We present the filtered black hole demographics and spin-mass distribution (Section \ref{subsec:bh_demographics}), comparing the SM16 environmental rate framework against the classical isothermal sphere baseline. We examine the cusp-core morphological divide and its effect on TDE budgets through radial profiles and running statistics (Section \ref{subsec:morph_divide}), and conclude with volumetric yields and 2D surface projections for all six simulated environments (Section \ref{subsec:tde_surface_proj}).

\subsection{Demographics and the relativistic efficiency correction}\label{subsec:bh_demographics}

\begin{figure}
    \centering
    \includegraphics[width=\linewidth]{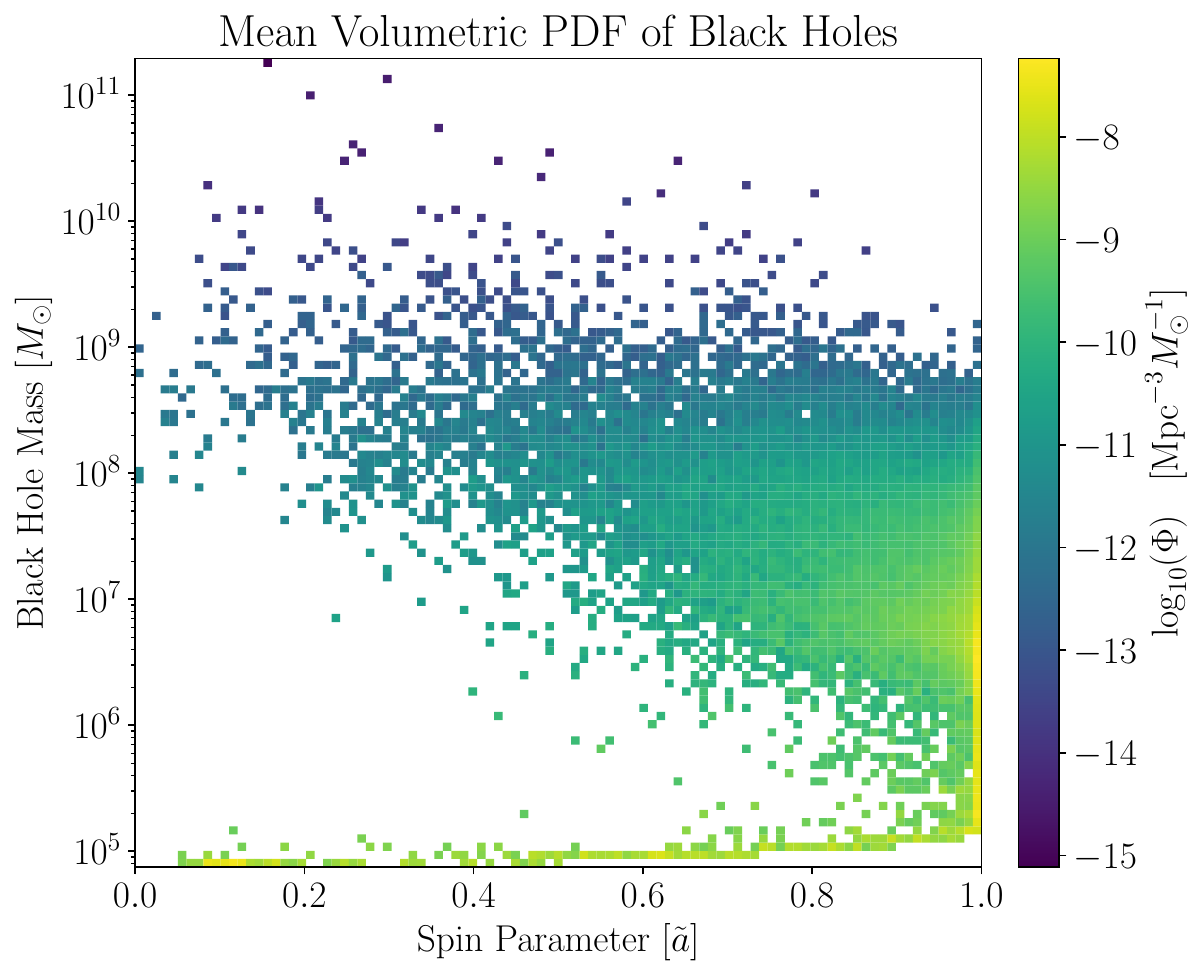}
    \caption{Combined volumetric probability density function of the filtered black hole population in the $\tilde{a}$-$\mbh$ parameter space across all simulated structures. To account for varying boundary sizes, the individual distributions were normalized by their respective zoom-in volumes before combination.}
    \label{fig:bh_population_map}
\end{figure}

Simulated SMBHs span various evolutionary stages from seeding to merger-driven growth. To properly interpret the underlying TDE rate, it is crucial to first establish the baseline spin-mass distribution of the simulated SMBHs after the main-sequence filters are applied. Figure \ref{fig:bh_population_map} shows the combined probability density function (PDF) of the black hole population in the $\atilde$-$\mbh$ parameter space. To prevent volumetric bias, the underlying distributions were normalized by their respective environmental volumes before combination. The distinct horizontal band of low-mass BHs ($\sim 10^5\,\msol$) spanning all spins is attributed to recently seeded black holes. As demonstrated by \cite{Sala2024}, this specific population consists of black holes that still show a signature of their initial seed conditions (i.e., zero spin), but rapidly evolve toward maximum spin while accreting 1-2 times their initial mass. Beyond this seed phase, there is a high concentration of low-mass ($10^5$-$10^7\,\msol$) BHs exhibiting maximum spin, and the figure demonstrates that the vast majority of the overall BH population is distributed in the high-spin regime.

To have a direct comparison to the TDE rate sensitivity to its surrounding stellar environment, as introduced in Section~\ref{subsec:vol_tde_rates}, we calculated $\gvol$ inferred from the standard baseline rate approximation from \cite{Kesden2012}:
\begin{equation}\label{eq:kesden_baserate}
    \Gamma_\mathrm{iso} = 4.8\times 10^{-4}\left(\frac{M}{10^6\,\msol}\right)^{-0.19}\,\yr^{-1}.
\end{equation}

$\Gamma_\mathrm{iso}$ assumes that the galactic center resembles the density profile of a single isothermal sphere \citep{Wang2004} and inherits the \cite{Schulze2011} $\mbh$-$\sigma$ relation. For each $\Gamma_\mathrm{iso}$, we applied the efficiency $\eta$ to account for direct captures and calculated the mean volumetric rate taken from all six zoom-in volumes, resulting in a high mean $\Gamma_\mathrm{vol, iso}=2860\,\mathrm{Gpc}^{-3}\,\yr^{-1}$.

\subsection{The morphological divide: cusps vs. cores}\label{subsec:morph_divide}

In contrast to assuming a density profile of an isothermal sphere, we applied the environmentally resolved \cite{Stone2016} base rates, introduced in Section \ref{subsec:vol_tde_rates}, using the actual density slopes $g$ taken from the simulation. For each BH, we applied identical $\eta$ efficiencies to $\Gamma_{\mathrm{vol},g}$ as previously applied to the standard baseline rate approximation. The index $g$ denotes the TDE rate sensitive to the local stellar density slope. The resulting mean volumetric TDE rate $\Gamma_{\mathrm{vol},g}\approx 606\,\mathrm{Gpc}^{-3}\,\yr^{-1}$ is significantly lower than $\Gamma_\mathrm{vol, iso}$.

This systematically higher TDE rate of $\Gamma_\mathrm{vol, iso}=2860\,\mathrm{Gpc}^{-3}\,\yr^{-1}$ stems from the crude assumption that the inner $1\,\kpc$ of all host galaxies behaves as a single isothermal sphere exhibiting pure steep cusps. However, in reality, galaxies show a strongly bimodal distribution of density slopes in their centers (see Figure \ref{fig:gamma_distribution}), indicating the existence of a massive population of core galaxies ($g\leq 1$). Because core galaxies have significantly longer relaxation times, they naturally produce fewer TDEs than cusps of the same mass. By applying the base rate from Equation \ref{eq:kesden_baserate}, we blindly assign high cusp-like rates to core galaxies, artificially inflating the total volumetric rate.

Figure \ref{fig:baserate_comparison} shows the theoretical mean TDE base rate as a function of black hole mass for the SM16 framework alongside the earlier analytical model of \cite{Kesden2012}. While this preceding formulation predicts a single high-yielding rate across all galaxies, the SM16 framework systematically lowers the expected flux. This reduction comes not only from explicit separation of cusp and core galaxies, but also from the adoption of a more realistic Kroupa present-day mass function \citep{Stone2016}. Furthermore, the framework utilizes the updated $\mbh$-$\sigma$ relationship established by \cite{McConnell2013}. After subdividing our sample by the measured density slopes $g$, the discrepancy is significantly amplified, since the vast majority of massive simulated galaxies fall into the core regime (see Figure \ref{fig:gamma_distribution}) and thus produce TDE rates roughly an order of magnitude lower than the isothermal assumption.

\begin{figure}
    \centering
    \includegraphics[width=\linewidth]{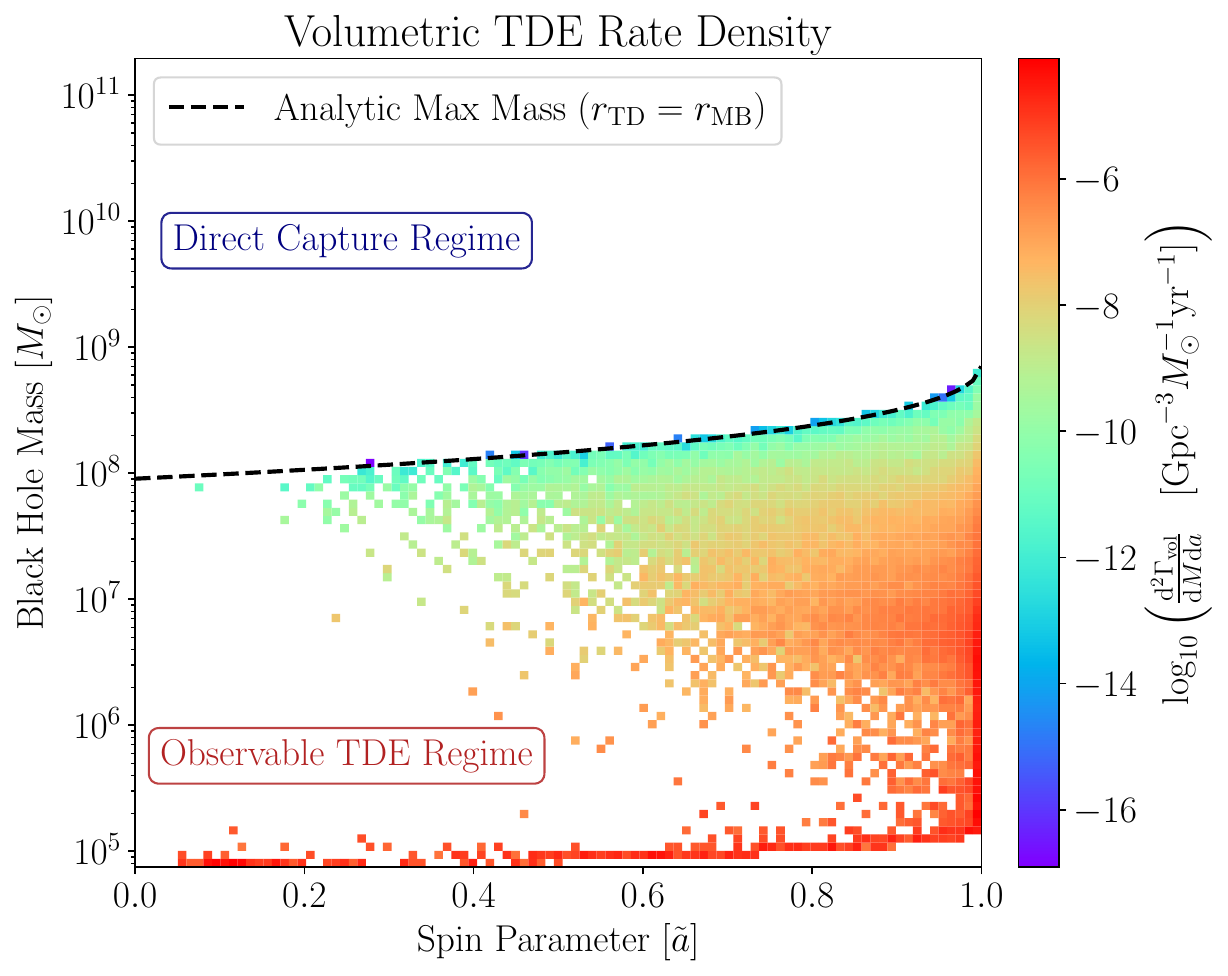}
    \caption{Mean volumetric TDE density taken from all six environments in this sample in $\tilde{a}$-$\mbh$ parameter space. The dashed line represents the analytical maximum black hole mass $M_{\bullet,\mathrm{max}}$ capable of producing a TDE.}
    \label{fig:tde_rate_density}
\end{figure}

Figure \ref{fig:tde_rate_density} shows the mean volumetric TDE rate density $\mathrm{d}^2\Gamma/\mathrm{d}M\mathrm{d}a$ in $\tilde{a}$-$\mbh$ parameter space taken from the full set of six simulated environments. It can be seen that higher rate densities follow a similar distribution as the probability density function of black holes in this parameter space (see Figure \ref{fig:bh_population_map}) because the volumetric rates are summed in each $\tilde{a}_i$-$M_{\bullet,i}$-segment. This results in higher TDE rate densities in high-spin, low-mass environments. In high $\mbh$-regimes $\eta\rightarrow 0$, resulting in a descending transition towards the analytical limit for black holes capable of visibly tidally disrupting stars. Recently seeded black holes in the $\sim 10^5\,\msol$ regime yield high TDE rate densities due to their large abundance and low mass, which independently produce high TDE rates regardless of spin.

Table \ref{tab:cluster_tde_rates} presents the resulting volumetric and absolute TDE rates across the different simulated environments. The most massive supercluster regions (Coma, Hercules, and Shapley) accumulate at a similar volumetric rate of roughly $761-794\,\gpc^{-3}\,\yr^{-1}$. This contrasts with the lower-mass environments, namely the Virgo supercluster region and the Fornax galaxy cluster, showing a rate of around $303-371\,\gpc^{-3}\,\yr^{-1}$. This significant environmental dependence underscores that $\gvol$ is not a universal constant. However, as we show below, this apparent mass correlation reflects differences in the black holes' concentration ($n_\bullet$) and population demographics rather than a direct dependence on the central anchor cluster mass.

\begin{table*}
    \centering
    \caption{Properties and TDE rates for the six simulated environments. Listed are the central anchor cluster mass $M_{500}$, the number of valid black holes within the volume $N_\bullet$, the global black hole number density $n_\bullet$, the 3D black hole half-number radius $R_{1/2, \bullet}$, the inscribed safe sphere radius $R_\mathrm{safe}$, the central anchor cluster radius $R_{500}$, the cusp to core fraction $f_\mathrm{cusp/core}$, the volumetric TDE rate $\Gamma_\mathrm{vol}$, the absolute TDE rate per zoom-in boundary volume $\Gamma_\mathrm{abs}$, the average TDE rate per black hole $\Gamma_\mathrm{norm}$, and the average TDE rate per black hole relative to the sample mean $\tilde{\Gamma}$.}
    \setlength{\tabcolsep}{5pt}
    \begin{tabular}{l c c c c c c c c c c c}
        \hline\hline 
        Name & $M_{500}$ & $N_\bullet$ & $n_\bullet$ & $R_{1/2, \bullet}$ & $R_\mathrm{safe}$ & $R_{500}$ & $f_\mathrm{cusp/core}$ & $\Gamma_\mathrm{vol}$ & $\Gamma_\mathrm{abs}$ & $\Gamma_\mathrm{norm}$ & $\tilde{\Gamma}$ \\
        & $[10^{14}\,M_\odot]$ & & $[\mathrm{Mpc}^{-3}]$ & $[\mathrm{Mpc}]$ & $[\mathrm{Mpc}]$ & $[\mathrm{Mpc}]$ & & $[\mathrm{Gpc}^{-3}\,\mathrm{yr}^{-1}]$ & $[\mathrm{yr}^{-1}]$ & $[10^{-5}\,\mathrm{yr}^{-1}]$ & \\
        \hline
        Coma & 17.24 & 10180 & $1.72 \times 10^{-2}$ & 29.49 & 22.29 & 2.76 & 0.98 & 777.65 & 0.46 & 4.52 & 1.00 \\
        Hercules & 13.98 & 4458 & $1.82 \times 10^{-2}$ & 19.76 & 20.12 & 2.57 & 1.00 & 794.22 & 0.20 & 4.38 & 0.97 \\
        Shapley & 13.87 & 13820 & $1.74 \times 10^{-2}$ & 31.01 & 31.54 & 2.57 & 0.90 & 761.24 & 0.60 & 4.38 & 0.97 \\
        Virgo & 9.05 & 2616 & $8.39 \times 10^{-3}$ & 20.81 & 23.36 & 2.23 & 0.83 & 370.74 & 0.12 & 4.42 & 0.98 \\
        Perseus & 9.02 & 5829 & $1.38 \times 10^{-2}$ & 31.03 & 18.73 & 2.22 & 0.98 & 626.58 & 0.27 & 4.55 & 1.01 \\
        Fornax & 0.35 & 2567 & $6.24 \times 10^{-3}$ & 31.26 & 8.45 & 0.75 & 0.99 & 302.79 & 0.12 & 4.85 & 1.07 \\
        \hline
    \end{tabular}
    \label{tab:cluster_tde_rates}
\end{table*}

We report an average TDE rate per black hole ($\Gamma_\mathrm{norm}=\gabs/N_\bullet$) across the full sample of $4.5\times 10^{-5}\,\yr^{-1}$, which is a highly stable metric across the six simulated environments (see Table \ref{tab:cluster_tde_rates}). Interestingly, the average TDE rate per black hole relative to the sample mean
\begin{equation}
 \tilde{\Gamma}=\frac{\Gamma_\mathrm{norm}}{\langle \Gamma_\mathrm{norm} \rangle}
\end{equation}
does neither scale with the volumetric nor with the absolute TDE rate. This implies that TDE rates do not depend solely on the TDE production efficiency of individual black holes within an environment. 

Instead, $\tilde{\Gamma}$ is primarily driven by the black hole mass distribution. Environments hosting a disproportionate excess of low-mass black holes avoiding the Hills mass cutoff appear as more efficient, independently of the volumetric rate. Figure \ref{fig:population_deviations} illustrates this by depicting the difference between the cumulative distribution function ($\Delta\mathrm{CDF}$) for individual environments and the ensemble average. Fornax emerges as a clear outlier, hosting disproportionately more low-mass black holes (left panel), meaning a significantly larger fraction of its black hole population safely avoids the direct capture constraints. Because this mass excess predominantly occurs in the low-mass regime where the restrictive Hills cutoff does not strongly apply, these black holes maintain highly efficient loss cones even at systematically lower spins (right panel, Figure \ref{fig:population_deviations}). Alongside Fornax, the Coma and Perseus supercluster environments show a disproportional excess of black hole masses below the Hills mass limit, being the two remaining environments with $\tilde{\Gamma}\geq 1$.

This trend cannot be explained by differences in the underlying morphological demographics of these structures, since the cusp-core fraction $f_\mathrm{cusp/core}$ does not directly imply a higher TDE rate in the zoom-in volumes. In fact, the combination of $N_\bullet$, $n_\bullet$, and $R_{1/2,\bullet}$ reveals that the spatial distribution and concentration of black holes have a significant influence on the TDE rates, and in consideration of $M_{500}$, links to the cluster's assembly can be made. Fornax with the highest $\tilde{\Gamma}$, serves as a good example, highlighting that, with low $n_\bullet$ and large $R_{1/2,\bullet}$, the black holes are widely distributed across the zoom-in volume. Therefore, fewer individual galaxies merged to form larger ones, resulting in a significantly higher number of low-mass black holes compared to the average of all six environments. 

A direct contrast to this efficiency-driven regime is observed in the Virgo supercluster region. While Virgo and Fornax host similar $N_\bullet$ and yield similar $\gabs$, Fornax is significantly more efficient on a per-black hole basis ($\Gamma_\mathrm{norm}$ and $\tilde{\Gamma}$). Even though Virgo contains a substantially lower fraction of cuspy galaxies than Fornax, and lacks Fornax's disproportionate excess of low-mass black holes in the $(10^6-10^8)\,\msol$ mass regime, it still produces higher $\gvol$. Virgo compensates for its lower per-black hole efficiency through its higher $n_\bullet$. Because black holes in Virgo are more densely packed than those in Fornax, a high $\gvol$ is recovered through spatial concentration, demonstrating that the macroscopic environmental density can overpower small-scale black hole properties to drive the volumetric TDE budget. 

\begin{figure}
    \centering
    \includegraphics[width=1\linewidth,trim=0cm 1.1cm 0cm 0cm]{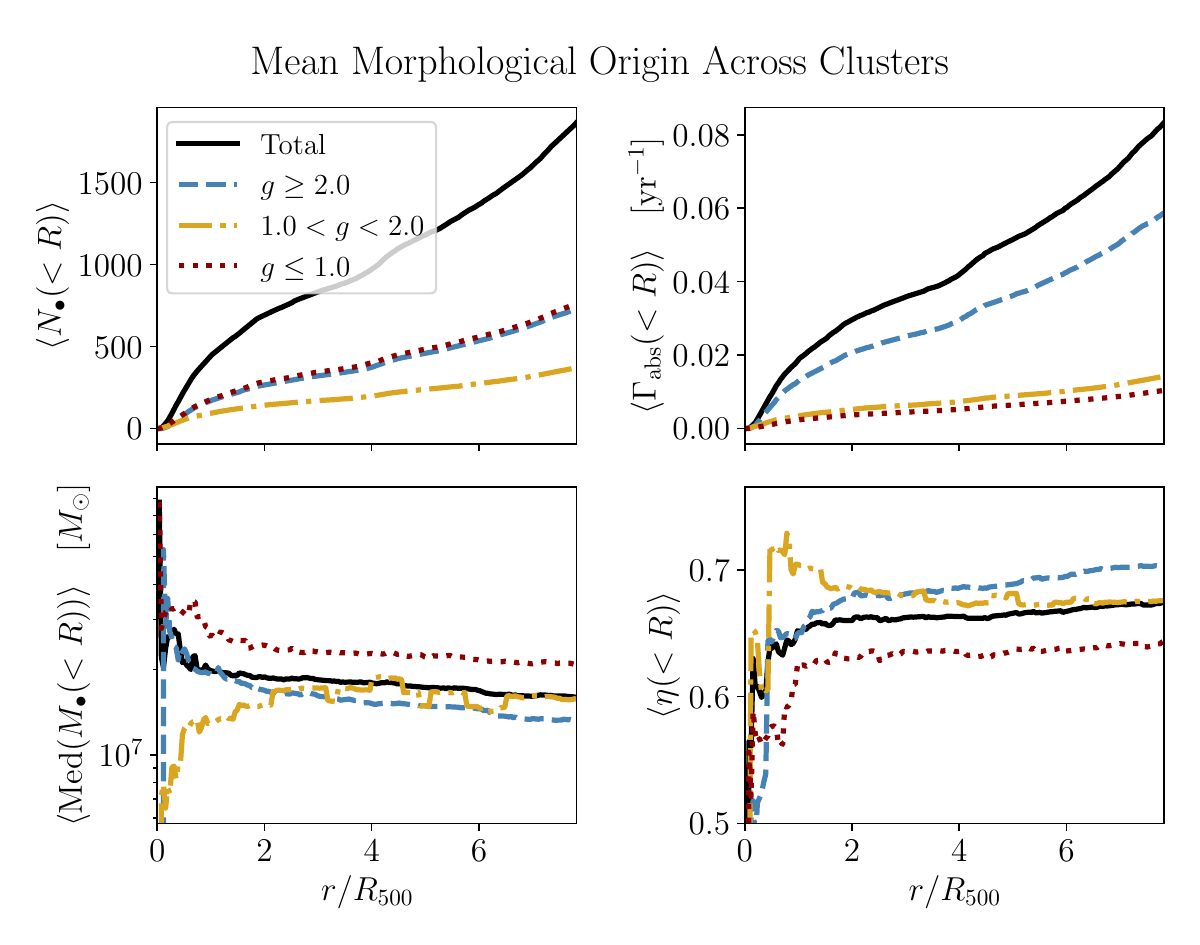}
    \caption{Mean cumulative morphological properties of black holes, averaged across all six simulated environments as a function of $r/R_{500}$, separated into cores ($g\leq 1$), transition ($1<g<2$), and cusps ($g\geq 2$). In the top row, the left panel shows the mean cumulative number of valid black holes ($\langle N_\bullet(<R)\rangle$), and the right panel depicts the mean cumulative absolute TDE rate ($\langle\gabs(<R)\rangle$). The bottom row displays the running median black hole mass of the enclosed population ($\langle\mathrm{med}(\mbh(<R))\rangle$) on the left and the running mean Kesden efficiency ($\langle\eta(<R)\rangle$) on the right. While core and cusp galaxies exist in similar numbers, the higher median black hole mass in cored galaxies reduces their Kesden efficiency as the Hills cutoff dominates, allowing cuspy satellite galaxies to dominate the environment's total transient yield.}
    \label{fig:mean_rad_tde_morphology}
\end{figure}

As shown in Figure \ref{fig:mean_rad_tde_morphology}, the mean radially cumulative number of BHs across all six simulated structures with cuspy stellar density profiles roughly matches the number of black holes in cored galaxies across all radii, while transition galaxies ($1<g<2$) represent a demographic minority. The top-right panel shows the mean radially cumulative TDE rate out of all six environments. The largest fraction of TDEs is produced by galaxies hosting cuspy stellar density profiles. One reason is that the SM16 framework predicts a higher baseline TDE rate for cuspy galaxies (see Figure \ref{fig:baserate_comparison}). However, the main reason that a larger fraction of TDEs is driven by cuspy galaxies is their greater internal efficiency in producing them.

The bottom-left panel in Figure \ref{fig:mean_rad_tde_morphology} shows the median black hole mass of the enclosed population averaged over all six zoom-in volumes. It reveals that black holes originating from core galaxies result in $\langle\mathrm{med}(\mbh(<R))\rangle$ twice as high as for cuspy environments up to maximum radii. Cored galaxies, therefore, push their black hole population more towards $M_{\bullet,\,\mathrm{max}}$ than cuspy galaxies. The bottom-right panel depicts the running Kesden efficiency. Cuspy galaxies significantly dominate $\langle\eta(<R)\rangle$ across all distances over cores. Since cored galaxies host significantly heavier black holes, the higher mass cutoff counteracts the spin extensions, driving $\eta$ down. Cuspy galaxies, on the other hand, benefit from high intrinsic loss-cone feeding and low-mass black holes that safely convert those disruptions into observable flares. Generally, strong TDE contributions of cuspy galaxies stem from the far outskirts of massive structures because the running median black hole mass drops at increasing radii as numerous low-mass satellites are enclosed, which in turn pushes $\eta$ to higher efficiencies. In conclusion, the environment's true yield is dominated by the higher baseline TDE rate in combination with higher $\eta$ coming from cuspy galaxies.

\subsection{Volumetric yields and 2D surface projections}\label{subsec:tde_surface_proj}

\begin{figure}
    \centering
    \includegraphics[width=\linewidth]{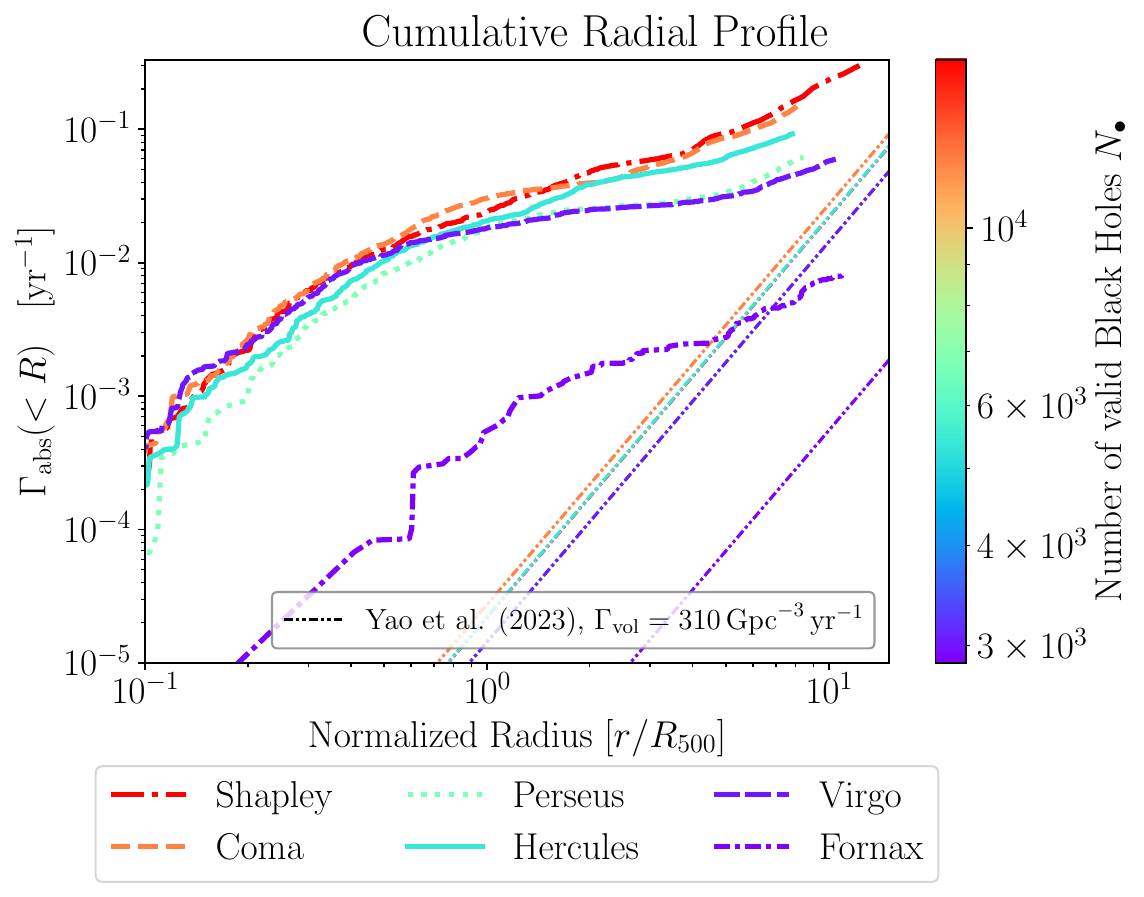}
    \caption{Cumulative absolute TDE rate ($\gabs$) as a function of the normalized cluster radius ($r/R_{500}$). The maximum radius is determined by $R_\mathrm{safe}$. The profiles are color-coded by the total number of valid black holes ($N_\bullet$) per environment. While inner regions show similar integrated TDE rates across all six structures, $N_\bullet$-rich environments show greater variation in the extended halo. The densely dash-dot-dotted lines, normalized by the individual $R_{500}$ of the underlying sample, represent simplified cumulative absolute TDE rates extrapolated from the empirical volumetric rate of $310\,\gpc^{-3}\,\yr^{-1}$ for events with $L_\mathrm{bb}>10^{43}\,\erg^{-1}\,\yr^{-1}$ \citep{Yao2023}, assuming a homogeneous spherical volume.}
    \label{fig:cum_rad_tde_rate}
\end{figure}

The simulation produces absolute TDE yields ranging from 0.12 to $0.60\,\yr^{-1}$ within zoom-in regions with radial extents of $55-92\,\mpc$. Environmental properties and SMBH abundance strongly impact the total TDE yield within a zoom-in region. Figure \ref{fig:cum_rad_tde_rate} shows the radial profiles of the cumulative absolute TDE rate $\gabs(<R)$ for the six zoom-in volumes. To further illustrate how strongly simulated TDE rates are influenced by spatial clustering rather than uniform volume filling, this figure includes empirical baselines for comparison. Each baseline color-coded to match its simulated counterpart is normalized by $R_{500}$ of the respective environment. These lines are obtained by extrapolating the constant volumetric TDE rate of $\Gamma_\mathrm{vol,Yao}=310\,\gpc^{-3}\,\yr^{-1}$ reported by \cite{Yao2023} into a homogeneous spherical volume ($\Gamma_\mathrm{abs,Yao}(<R)=4/3\pi R^3 \,\Gamma_\mathrm{vol,Yao}$). Comparing the simulated profiles to this uniform baseline highlights that TDE yields do not grow linearly with volume. Color-coded by $N_\bullet$, it is clear that the more black holes are enclosed in this volume, the higher the total rate can be achieved. Table \ref{tab:cluster_tde_rates} shows that the Hercules supercluster region is more massive than Shapley and Perseus but hosts significantly fewer black holes than these two structures. Furthermore, Hercules exhibits the highest $\gvol$ out of all environments, because it is a highly concentrated region with the largest global black hole number density $n_\bullet$ and the smallest black hole half-number radius $R_{1/2,\bullet}$. Even though Hercules hosts less than half as many black holes as compared to Coma and Shapley, it appears to be one of the most promising TDE hosts by means of $\gvol$, even though $\tilde{\Gamma}$ is one of the lowest of all.

This conundrum is partially resolved in Figure \ref{fig:radial_comparison_hercules}; the $3\times3$-panel figure shows the cumulative radial profiles of the median black hole mass $\mathrm{Med}(M_\bullet(<R))$, the dimensionless spin $\tilde{a}(<R)$, and the Kesden efficiency $\eta(<R)$ of Hercules compared to the rest of the sample for galactic cusp, core, and transition environments. While the dashed line represents the median, the transparent band is defined by the 10th-90th percentile range across all five other volumes, excluding Hercules. Especially in cuspy and cored environments, black holes in Hercules are significantly more massive and possess a lower spin than the rest of the sample, leading to a systematic reduction of $\eta$. Generally, a lower $\eta$ directly results in a reduction of $\tilde{\Gamma}$, because more black holes compared to other environments directly capture stars without visibly tidally disrupting them.

\begin{figure}
    \centering
    \includegraphics[width=1\linewidth,trim=0cm 0.5cm 0cm 0cm]{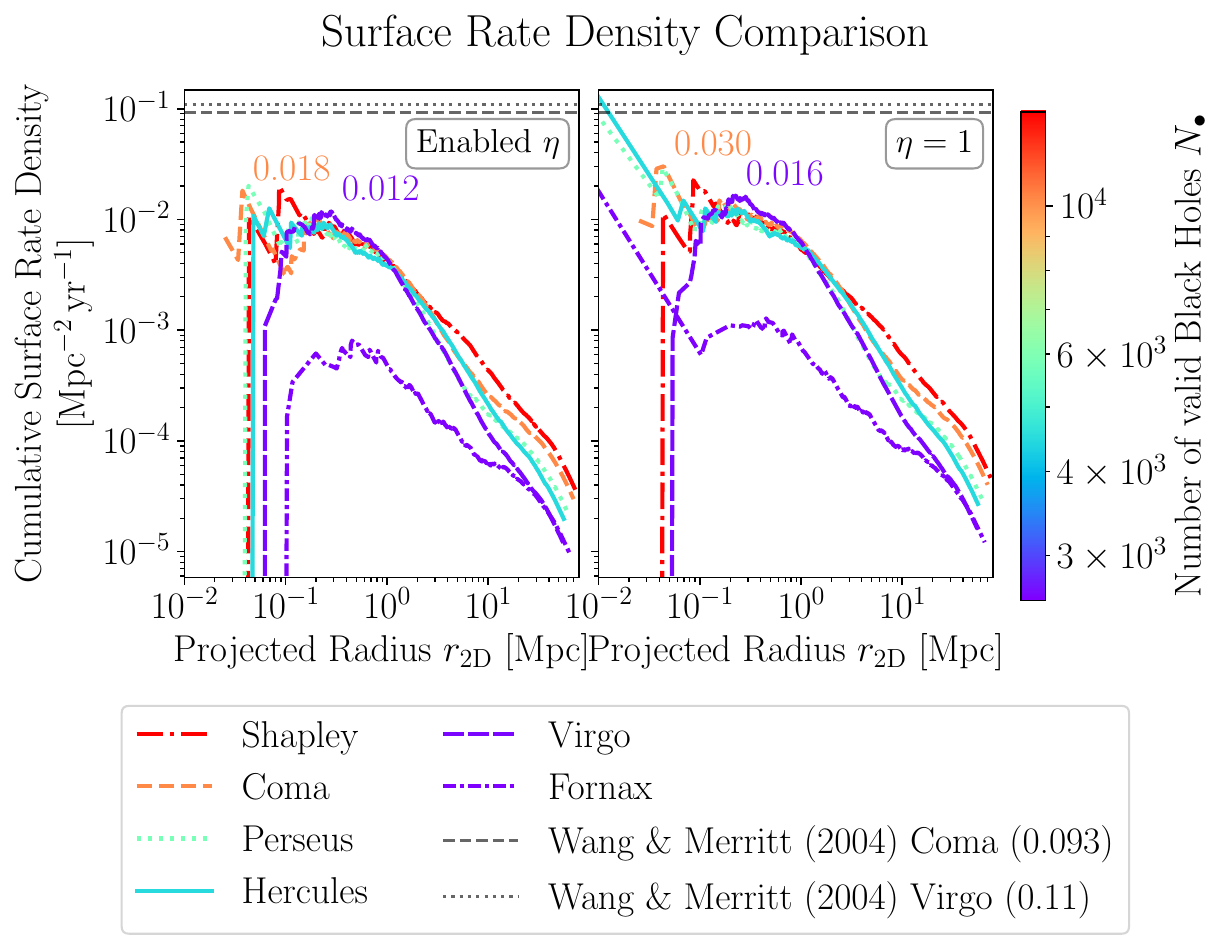}
    \caption{Cumulative surface rate density produced by a 2D surface projection as a function of the projected radius. The two-panel figure compares the rates for true Kesden efficiencies $\eta$ and $\eta=1$. The horizontal lines mark the surface rate density literature values by \cite{Wang2004}. In both panels, it can be seen that the simulated profiles underestimate the reported values for Coma ($0.093\,\mpc^{-2}\,\yr^{-1}$) and Virgo ($0.11\,\mpc^{-2}\,\yr^{-1}$) approximately by one order of magnitude. The profiles, color-coded by the number of black holes $N_\bullet$, show that the peak rate densities do not depend on $N_\bullet$.}
    \label{fig:cum_tde_surf_rate_dens_comp_kesden}
\end{figure}

Figure \ref{fig:cum_tde_surf_rate_dens_comp_kesden} presents a 2D surface projection depicting the cumulative surface rate density. To provide a direct comparison of the influence of the Kesden efficiency parameter $\eta$, the figure shows the surface rate density with true $\eta$ values and with $\eta=1$ only. Both peak at approximately $10^{-2}\,\mpc^{-2}\,\yr^{-1}$ with minimum differences in peak values reached. Since the surface rate densities are normalized by the respective enclosed areas, the radial profiles show a general similarity in shape and magnitude, except for Fornax. This outlier can be explained by a combination of parameters from Table \ref{tab:cluster_tde_rates}. Even though Fornax is a highly efficient TDE factory (\cf{} $\tilde{\Gamma}$) and exhibits a $f_\mathrm{cusp/core}$ higher than average, while having $\gabs$ and $\gvol$ similar to Virgo, the low surface rate density yield stems from lowest $n_\bullet$ while simultaneously having the highest $R_{1/2,\bullet}$ out of all six environments. Presumably, the low $M_{500}$ drives the wide spatial spread of the sparse number density of black holes in comparison to other structures due to a low gravitational potential.

\section{Discussion}\label{sec:discussion}

We reconcile our simulated rates with observational constraints (Section \ref{subsec:theory_vs_obs}) and demonstrate how traditional 2D extrapolations overestimate central surface densities, while the true budget is driven by peripheral dwarf galaxies. We then decompose the observed discrepancy into dynamical core depletion and relativistic extensions to the Hills mass cutoff (Section \ref{subsec:relativistic_and_core_depl}), and place our findings in a cosmological context, showing how actively assembling superclusters systematically suppresses per-BH efficiency (Section \ref{subsec:cosmo_context}).

\subsection{Reconciling theoretical rates to observational constraints}\label{subsec:theory_vs_obs}

Our simulated per black hole TDE rate of $4.5\times 10^{-5}\,\yr^{-1}$ (where each black hole in the simulation hosts its own galaxy/subhalo) sits at the upper end of current observational constraints. X-ray all-sky surveys by \cite{Sazonov2021} and \cite{Grotova2025} report rates of $(1.1\pm 0.5)\times 10^{-5}\,\pyrpgal$ and $\approx 1.2\times 10^{-5}\,\pyrpgal$, respectively, 
where the latter explicitly characterize their sample as the \enquote{unobscured thermal population} of TDEs, implying that obscured events are systematically excluded. Mid-IR observations by \cite{Masterson2024}, which probe dust-obscured TDEs missed in optical bands, place a higher lower limit of $(2.0\pm 0.3)\times 10^{-5}\,\pyrpgal$. Optical surveys from ZTF find rates of $\sim 3.2\times 10^{-5}\,\pyrpgal$ \citep{Yao2023}. These multi-wavelength constraints highlight the impact of observational biases: X-ray rates represent a strict lower bound due to severe nuclear obscuration, while optical and mid-IR surveys recover larger fractions of the true population. Our simulated rate of $4.5\times 10^{-5}\,\yr^{-1}$ sits slightly above these yields, reflecting the complete theoretical budget prior to the application of survey-specific selection functions or viewing-angle attenuation. Still, both simulation and observation agree that the true rate lies well below the $\gtrsim 10^{-4}\,\pyrpgal$ predicted for low-mass cuspy galaxies by idealized two-body relaxation models \citep[e.g.][]{Magorrian1999, Wang2004}.

In the simulation, the absolute TDE yield ($0.12\,\yr^{-1}$) produced by Virgo is in remarkable agreement with the pioneering calculations of \cite{Wang2004}, who predicted an absolute rate of approximately $0.16\,\yr^{-1}$. These absolute rates were achieved by taking a 2D central surface rate ($0.11\,\mpc^{-2}\,\yr^{-1}$ for Virgo, and $0.093\,\mpc^{-2}\,\yr^{-1}$ for Coma) and integrating an assumed 2D spatial distribution of dwarf galaxies with a steep exponential drop-off ($R_0 = 0.48\,\mpc$). This steep exponential fit forces a large population of intact dwarf galaxies to reside within the innermost megaparsec. However, when comparing the simulated cumulative surface TDE rate density ($\Sigma_{r<R}\Gamma/(\pi R^2)$) to the literature values, Figure \ref{fig:cum_tde_surf_rate_dens_comp_kesden} shows the 2D surface rate density, mimicking the line-of-sight projection used in the observational literature, it highlights that the population of projected surface event rates in Coma and Virgo are overestimated by the literature. This has the consequence that extrapolating the projected 2D central density squeezes the entire TDE budget into the cluster core, leading to overall similar absolute TDE rates between the simulation and \cite{Wang2004} for Virgo. The true simulated central surface rates, on the other hand, are reaching peaks that are a magnitude lower, around $10^{-2}\,\mpc^{-2}\,\yr^{-1}$. This figure further shows that this discrepancy cannot be resolved by introducing the tidal disruption efficiency $\eta$.

\subsection{Dynamical core depletion and relativistic corrections}\label{subsec:relativistic_and_core_depl}

This discrepancy is influenced by two factors: the relativistic efficiency correction and the dynamical depletion of low-mass subhalos. The first minor, but non-negligible effect is that \cite{Wang2004} pre-dates the formal quantification of the correction established by \cite{Kesden2012}. An incorporation of this relativistic effect lowers the TDE rates of the massive black holes hosted by the central cored galaxies. This effect is visually evident in Figure \ref{fig:tde_rate_density}, which shows the volumetric TDE rate density in the $\tilde{a}$-$\mbh$ parameter space. In the high-mass regime, the rate density drops sharply as $\eta\rightarrow 0$ while approaching the analytical Hills limit where direct capture dominates over observable disruption. This theoretical suppression is observationally confirmed: \cite{Nair2026} identify a luminosity function break in mid-IR selected TDEs at $\log(L_\mathrm{peak,\,W2}/\mathrm{erg}\,\s^{-1})\simeq 43.4$, where high-luminosity events are suppressed by more than 200 times. Based on the findings that the peak IR luminosity scales with the black hole mass \citep{Wang2018}, \cite{Nair2026} conclude that their break luminosity aligns with the Hills mass cutoff ($\mbh\approx 10^8\,M_\odot$), independently validating that massive black holes systematically fail to produce observable flares. However, the relativistic Hills limit is not the reason why the observed surface rate densities for Coma and Virgo are by an order of magnitude higher. The dominant driver is dynamical core depletion, in which low-mass subhalos merge into larger ones in the extreme inner core, in contrast to the assumed high surface density of dwarf galaxies. \cite{Secker1996} observationally validates the hypothesis that faint dwarf ellipticals are destroyed in the dense cluster cores.

\subsection{Cosmological context and theoretical upper limits}\label{subsec:cosmo_context}

Yet, by resolving the full zoom-in region up to the cluster outskirts, the simulation recovers the high absolute yield for Virgo and suggests that Coma actually produces roughly $3.8$ times more TDEs than Virgo, because the yield is integrated over the full zoom-in volume that analytical models assumed was empty. Especially at radii far outside $R_{500}$ of a galaxy cluster, the cumulative TDE rate is driven by the number of galaxies falling towards the central potential. In Figure \ref{fig:cum_rad_tde_rate}, it can be seen that Coma progressively recovers its high absolute TDE rate at distances of several $R_{500}$. High TDE rates from the far extents of galaxy clusters can be recovered because their black hole mass is statistically lower, resulting in higher $\eta$ (Figure \ref{fig:cum_rad_tde_rate}).

\cite{Zoeller2025} showed that while the zoom-in simulation reproduces the Coma cluster in observations from the Wendelstein $2.1\,\mathrm{m}$ telescope, its subgrid physics may overproduce intermediate- to low-luminosity dwarf galaxies, where this effect appears considerably weaker when rescaling the simulation. Since the simulated absolute TDE budget is driven by this peripheral dwarf population, the derived volumetric yields should be treated as a robust theoretical upper limit. Furthermore, the massive $1.5\,\mathrm{deg}^2$ footprint observed by \cite{Zoeller2025} emphasizes that accounting for the transient-producing dwarf galaxies requires a large spatial coverage where a simple 2D core extrapolation would fail. As already indicated in Table \ref{tab:cluster_tde_rates}, this rate does not directly scale with the total cluster mass. \cite{Secker1996} highlight this by showing that Coma's dwarf-to-giant ratio is identical to Virgo's. Instead, TDE yields rather trace complex cosmic substructure and the absolute number of low-mass black holes per galaxy cluster.

Regarding the puzzling controversy in the Hercules supercluster environment, the low absolute number of black holes and the low $\eta$ values, in combination with the high concentration and the high cluster mass, can be a direct consequence of an evolved dynamical state of the cluster environment. Given their assembly history, black holes must have merged, resulting in environments that are less efficient at producing TDEs than in the remaining zoom-in volumes. The dense cluster environment compensates for the low $\tilde{\Gamma}$ by the high concentration of black holes, even though the absolute number of black holes is relatively low. This recovers a high $\gvol$, even though $\tilde{\Gamma}$ is one of the lowest. Observations confirm that Hercules is not a single, mature cluster, but a highly complex, unrelaxed supercluster. \cite{Monteiro-Oliveira2022} found that its core is composed of A2147, A2151, and A2152, which are themselves highly substructured, with A2151 alone consisting of 5 subclusters. \cite{Seidel2025} found that the simulated Hercules cluster (A2147) counterpart is not only fed by the individual clusters A2151 and A2152, but also by at least three massive filaments along which the subclusters are plunging inward, highlighting that the simulated Hercules environment is actively assembling at $z=0$. Furthermore, Hercules is one of the late-time-growing superclusters in the simulation, and it is one of the superclusters that will experience its final major merger earlier than Centaurus, Coma, Shapley, and Virgo \citep[\cf{} Figure 8,][]{Seidel2025}.

This continuous, violent hierarchical merging of clusters and other substructures along filaments provides an environment with more enhanced galaxy interactions than in mature, isolated clusters. Because of this ongoing group-cluster interaction and the dense spatial packing (as can be seen in $n_\bullet$ and $R_{1/2,\bullet}$), galaxy-galaxy mergers are naturally accelerated within the subhalos. This systematically drives up the overall black hole masses across the entire population of galaxies, even the ones retaining cuspy profiles (see Figure \ref{fig:radial_comparison_hercules}). Simultaneously, this violent environment likely explains the low absolute number of black holes within this volume. This interconnection between cluster dynamics and TDE yields highlight an explicit link between the macroscopic cosmic web and microscopic loss-cone dynamics. Actively assembling, unrelaxed supercluster environments, such as Hercules and Shapley, inherently reduce localized TDE production efficiency due to merger-driven black hole growth, while preserving high volumetric yields (see Table \ref{tab:cluster_tde_rates}).

\section{Conclusion}

This study bridges the gap between theoretical predictions and observed TDE rates by examining simulated large-scale zoom-in regions while modeling the local stellar environment around black holes. This approach allows us to properly analyze the loss-cone dynamics on sub-galactic scales within a fully 3D cosmological framework.

We demonstrate that volumetric ($\gvol$) and per-black hole ($\Gamma_\mathrm{norm}$) TDE rates do not necessarily correlate within the same simulated environment. The macroscopic black hole number density ($n_\bullet$) and the abundance of peripheral galaxies strongly alter the total TDE budget. For instance, the highly evolved supercluster Hercules exhibits high $\gvol$, even though $\Gamma_\mathrm{norm}$ is low due to merger-driven black hole mass growth, which drives the Kesden efficiency toward zero. On the other hand, a high $\Gamma_\mathrm{norm}$ can be reached in clusters where $\gvol$ is low. Fornax, for example, hosts the most efficient cluster environment in producing TDEs, while maintaining the lowest $n_\bullet$ and $R_{1/2,\bullet}$. Thus, the environment with the lowest galaxy density is the most efficient at producing TDEs, implying that most galaxies in Fornax have not yet merged with others. This is consistent with Fornax being a merging low-mass cluster, with the Fornax A group still actively assembling in the simulation. This high efficiency is driven by an excess of low-mass, unmerged black holes that completely bypass the relativistic Hills cutoff. These low-mass black holes dominate the volumetric TDE rate density across the full spin range, confirming that the TDE budget is significantly influenced by the abundance of low-mass, unmerged systems rather than the per-event efficiency of the massive black holes. Clusters 
less efficient in producing TDEs as Fornax can still compensate for this effect by their higher number density of black holes, as demonstrated by Virgo. This underscores that global cluster properties fundamentally dictate local TDE rates.

Despite these vast environmental differences across both small and large scales, the simulated per-black hole TDE rates remain remarkably stable across all evaluated environments (Table \ref{tab:cluster_tde_rates}). Averaged across the full sample, we report a simulated per-black-hole TDE rate of $4.5\times 10^{-5}\,\yr^{-1}$. This result is in strong agreement with the current observational consensus of $\sim 3\times 10^{-5}\,\yr^{-1}$. Because the subgrid physics of the underlying cosmological model slightly overproduces the low-mass dwarf galaxies that dominate the absolute TDE budget, our derived rate of $4.5\times 10^{-5}\,\yr^{-1}$ serves as a robust theoretical upper limit for upcoming transient surveys.

\begin{acknowledgements}

JSS, IK, and KD acknowledge support by the COMPLEX project from the European Research Council (ERC) under the European Union’s Horizon 2020 research and innovation program grant agreement ERC-2019-AdG 882679. IK was supported by the Simons Foundation via the Simons Investigator Award to A. A. Schekochihin. LS acknowledges support by the Deutsche Forschungsgemeinschaft (DFG, German Research Foundation) under Germany’s Excellence Strategy - EXC-2094 - 390783311 and the Computational Center for Particle and Astrophysics (C2PAP). BS acknowledges support by the grant agreements ANR-21-CE31-0019 / 490702358 from the French Agence Nationale de la Recherche / DFG for the LOCALIZATION project. The calculations for the hydrodynamical simulations were carried out at the Leibniz Supercomputer Center (LRZ) under the projects pn68na and pn82ba. This work was supported by the French government under the France 2030 program with the reference ANR-21-IDES-0006. The Métropole Européenne de Lille and the University of Lille are also gratefully acknowledged for the funding and support granted to the WILL-CHAIRES-25-009-UNIVERSITWINS project.

\end{acknowledgements}



\bibliographystyle{aa} 
\bibliography{literature.bib} 

@ARTICLE{sorce18,
     author = {{Sorce}, Jenny G.},
      title = "{Galaxy clusters in local Universe simulations without density constraints: a long uphill struggle}",
    journal = {Monthly Notices of the Royal Astronomical Society},
       year = 2018,
      month = aug,
     volume = {478},
     number = {4},
      pages = {5199-5208},
        doi = {10.1093/mnras/sty1631},
 archivePrefix = {arXiv},
     eprint = {1806.09633},
 primaryClass = {astro-ph.CO},
     adsurl = {https://ui.adsabs.harvard.edu/abs/2018MNRAS.478.5199S}
 }

@ARTICLE{2015JHEAp...7..148K,
       author = {{Komossa}, S.},
        title = "{Tidal disruption of stars by supermassive black holes: Status of observations}",
      journal = {JHEAp},
         year = 2015,
        month = sep,
       volume = {7},
        pages = {148-157},
          doi = {10.1016/j.jheap.2015.04.006},
archivePrefix = {arXiv},
       eprint = {1505.01093},
 primaryClass = {astro-ph.HE},
       adsurl = {https://ui.adsabs.harvard.edu/abs/2015JHEAp...7..148K}
}

@ARTICLE{sorce+21,
     author = {{Sorce}, Jenny G. and {Dubois}, Yohan and {Blaizot}, J{\'e}r{\'e}my and {McGee}, Sean L. and {Yepes}, Gustavo and {Knebe}, Alexander},
      title = "{I - A hydrodynamical CLONE of the Virgo cluster of galaxies to confirm observationally driven formation scenarios}",
    journal = {Monthly Notices of the Royal Astronomical Society},
       year = 2021,
      month = jun,
     volume = {504},
     number = {2},
      pages = {2998-3012},
        doi = {10.1093/mnras/stab1021},
 archivePrefix = {arXiv},
     eprint = {2104.13389},
 primaryClass = {astro-ph.CO},
     adsurl = {https://ui.adsabs.harvard.edu/abs/2021MNRAS.504.2998S}
 }

@ARTICLE{1979SvAL....5...16L,
       author = {{Lidskii}, V.~V. and {Ozernoi}, L.~M.},
        title = "{Tidal triggering of stellar flares by a massive black hole}",
      journal = {SvAL},
         year = 1979,
        month = jan,
       volume = {5},
        pages = {16-19},
       adsurl = {https://ui.adsabs.harvard.edu/abs/1979SvAL....5...16L}
}

@ARTICLE{1981A&A....95...39G,
       author = {{Gurzadian}, V.~G. and {Ozernoi}, L.~M.},
        title = "{Accretion of the cloud of gas debris of stars disrupted by the tidal forces of a supermassive black hole}",
      journal = {A\&A},
         year = 1981,
        month = feb,
       volume = {95},
       number = {1},
        pages = {39-45},
       adsurl = {https://ui.adsabs.harvard.edu/abs/1981A&A....95...39G}
}

@INPROCEEDINGS{1989IAUS..136..543P,
       author = {{Phinney}, E.~S.},
        title = "{Manifestations of a Massive Black Hole in the Galactic Center}",
    booktitle = {The Center of the Galaxy},
         year = 1989,
       editor = {{Morris}, Mark},
       series = {IAU Symposium},
       volume = {136},
        month = jan,
        pages = {543},
       adsurl = {https://ui.adsabs.harvard.edu/abs/1989IAUS..136..543P}
}

@ARTICLE{sorce+24,
     author = {{Sorce}, Jenny G. and {Mohayaee}, Roya and {Aghanim}, Nabila and {Dolag}, Klaus and {Malavasi}, Nicola},
      title = "{Distortions of the Hubble diagram: Line-of-sight signatures of local galaxy clusters}",
    journal = {Astronomy and Astrophysics},
       year = 2024,
      month = jul,
     volume = {687},
        eid = {A85},
      pages = {A85},
        doi = {10.1051/0004-6361/202349073},
 archivePrefix = {arXiv},
     eprint = {2301.01305},
 primaryClass = {astro-ph.CO},
     adsurl = {https://ui.adsabs.harvard.edu/abs/2024A&A...687A..85S}
 }

@article{hernandezmartinez+24,
 title = {Simulating the {{LOcal Web}} ({{SLOW}}). {{II}}. {{Properties}} of Local Galaxy Clusters},
 author = {Hernández-Martínez, Elena and Dolag, Klaus and Seidel, Benjamin and Sorce, Jenny G. and Aghanim, Nabila and Pilipenko, Sergey and Gottlöber, Stefan and Lebeau, Théo and Valentini, Milena},
 date = {2024-07-01},
 year = {2024},
 journal = {Astronomy and Astrophysics},
 volume = {687},
 pages = {A253},
 publisher = {EDP},
 issn = {0004-6361},
 doi = {10.1051/0004-6361/202449460},
 url = {https://ui.adsabs.harvard.edu/abs/2024A&A...687A.253H},
 urldate = {2024-12-04}
 }

@ARTICLE{Damiano2024,
       author = {{Damiano}, Alice and {Valentini}, Milena and {Borgani}, Stefano and {Tornatore}, Luca and {Murante}, Giuseppe and {Ragagnin}, Antonio and {Ragone-Figueroa}, Cinthia and {Dolag}, Klaus},
        title = "{Dynamical friction and the evolution of black holes in cosmological simulations: A new implementation in OpenGadget3}",
      journal = {\aap},
         year = 2024,
        month = dec,
       volume = {692},
          eid = {A81},
        pages = {A81},
          doi = {10.1051/0004-6361/202450021},
archivePrefix = {arXiv},
       eprint = {2403.12600},
 primaryClass = {astro-ph.CO},
       adsurl = {https://ui.adsabs.harvard.edu/abs/2024A&A...692A..81D}
}

@ARTICLE{Damiano2025,
       author = {{Damiano}, Alice and {Borgani}, Stefano and {Valentini}, Milena and {Murante}, Giuseppe and {Tornatore}, Luca and {Strakos}, Petr and {Jaros}, Milan},
        title = "{Dynamical friction and massive black hole orbits: Analytical predictions and numerical solutions}",
      journal = {\aap},
         year = 2025,
        month = dec,
       volume = {704},
          eid = {A83},
        pages = {A83},
          doi = {10.1051/0004-6361/202556054},
archivePrefix = {arXiv},
       eprint = {2506.20740},
 primaryClass = {astro-ph.GA},
       adsurl = {https://ui.adsabs.harvard.edu/abs/2025A&A...704A..83D}
}

@ARTICLE{2016MNRAS.463.1797D,
       author = {{Dolag}, K. and {Komatsu}, E. and {Sunyaev}, R.},
        title = "{SZ effects in the Magneticum Pathfinder simulation: comparison with the Planck, SPT, and ACT results}",
      journal = {\mnras},
         year = 2016,
        month = dec,
       volume = {463},
       number = {2},
        pages = {1797-1811},
          doi = {10.1093/mnras/stw2035},
archivePrefix = {arXiv},
       eprint = {1509.05134},
 primaryClass = {astro-ph.CO},
       adsurl = {https://ui.adsabs.harvard.edu/abs/2016MNRAS.463.1797D}
}

@article{Dehnen2012,
    author = {Dehnen, Walter and Aly, Hossam},
    title = {Improving convergence in smoothed particle hydrodynamics simulations without pairing instability},
    journal = {Monthly Notices of the Royal Astronomical Society},
    volume = {425},
    number = {2},
    pages = {1068-1082},
    year = {2012},
    month = {09},
    issn = {0035-8711},
    doi = {10.1111/j.1365-2966.2012.21439.x},
    url = {https://doi.org/10.1111/j.1365-2966.2012.21439.x},
    eprint = {https://academic.oup.com/mnras/article-pdf/425/2/1068/4013442/425-2-1068.pdf},
}

@ARTICLE{Beck2016,
       author = {{Beck}, A.~M. and {Murante}, G. and {Arth}, A. and {Remus}, R.-S. and {Teklu}, A.~F. and {Donnert}, J.~M.~F. and {Planelles}, S. and {Beck}, M.~C. and {F{\"o}rster}, P. and {Imgrund}, M. and {Dolag}, K. and {Borgani}, S.},
        title = "{An improved SPH scheme for cosmological simulations}",
      journal = {\mnras},
         year = 2016,
        month = jan,
       volume = {455},
       number = {2},
        pages = {2110-2130},
          doi = {10.1093/mnras/stv2443},
archivePrefix = {arXiv},
       eprint = {1502.07358},
 primaryClass = {astro-ph.CO},
       adsurl = {https://ui.adsabs.harvard.edu/abs/2016MNRAS.455.2110B}
}

@ARTICLE{SpringelDiMatteo2006,
       author = {{Springel}, Volker and {Di Matteo}, Tiziana and {Hernquist}, Lars},
        title = "{Modelling feedback from stars and black holes in galaxy mergers}",
      journal = {\mnras},
         year = 2005,
        month = aug,
       volume = {361},
       number = {3},
        pages = {776-794},
          doi = {10.1111/j.1365-2966.2005.09238.x},
archivePrefix = {arXiv},
       eprint = {astro-ph/0411108},
 primaryClass = {astro-ph},
       adsurl = {https://ui.adsabs.harvard.edu/abs/2005MNRAS.361..776S}
}

@ARTICLE{Tornatore2007,
       author = {{Tornatore}, L. and {Borgani}, S. and {Dolag}, K. and {Matteucci}, F.},
        title = "{Chemical enrichment of galaxy clusters from hydrodynamical simulations}",
      journal = {\mnras},
         year = 2007,
        month = dec,
       volume = {382},
       number = {3},
        pages = {1050-1072},
          doi = {10.1111/j.1365-2966.2007.12070.x},
archivePrefix = {arXiv},
       eprint = {0705.1921},
 primaryClass = {astro-ph},
       adsurl = {https://ui.adsabs.harvard.edu/abs/2007MNRAS.382.1050T}
}

@ARTICLE{Tornatore2010,
       author = {{Tornatore}, L. and {Borgani}, S. and {Viel}, M. and {Springel}, V.},
        title = "{The impact of feedback on the low-redshift intergalactic medium}",
      journal = {\mnras},
         year = 2010,
        month = mar,
       volume = {402},
       number = {3},
        pages = {1911-1926},
          doi = {10.1111/j.1365-2966.2009.16025.x},
archivePrefix = {arXiv},
       eprint = {0911.0699},
 primaryClass = {astro-ph.GA},
       adsurl = {https://ui.adsabs.harvard.edu/abs/2010MNRAS.402.1911T}
}

@ARTICLE{Wiersma2009,
       author = {{Wiersma}, Robert P.~C. and {Schaye}, Joop and {Theuns}, Tom and {Dalla Vecchia}, Claudio and {Tornatore}, Luca},
        title = "{Chemical enrichment in cosmological, smoothed particle hydrodynamics simulations}",
      journal = {\mnras},
         year = 2009,
        month = oct,
       volume = {399},
       number = {2},
        pages = {574-600},
          doi = {10.1111/j.1365-2966.2009.15331.x},
archivePrefix = {arXiv},
       eprint = {0902.1535},
 primaryClass = {astro-ph.CO},
       adsurl = {https://ui.adsabs.harvard.edu/abs/2009MNRAS.399..574W}
}

@ARTICLE{Tornatore2004,
       author = {{Tornatore}, L. and {Borgani}, S. and {Matteucci}, F. and {Recchi}, S. and {Tozzi}, P.},
        title = "{Simulating the metal enrichment of the intracluster medium}",
      journal = {\mnras},
         year = 2004,
        month = mar,
       volume = {349},
       number = {1},
        pages = {L19-L24},
          doi = {10.1111/j.1365-2966.2004.07689.x},
archivePrefix = {arXiv},
       eprint = {astro-ph/0401576},
 primaryClass = {astro-ph},
       adsurl = {https://ui.adsabs.harvard.edu/abs/2004MNRAS.349L..19T}
}

@ARTICLE{Fabjan2010,
       author = {{Fabjan}, D. and {Borgani}, S. and {Tornatore}, L. and {Saro}, A. and {Murante}, G. and {Dolag}, K.},
        title = "{Simulating the effect of active galactic nuclei feedback on the metal enrichment of galaxy clusters}",
      journal = {\mnras},
         year = 2010,
        month = jan,
       volume = {401},
       number = {3},
        pages = {1670-1690},
          doi = {10.1111/j.1365-2966.2009.15794.x},
archivePrefix = {arXiv},
       eprint = {0909.0664},
 primaryClass = {astro-ph.CO},
       adsurl = {https://ui.adsabs.harvard.edu/abs/2010MNRAS.401.1670F}
}

@ARTICLE{Hirschmann2014,
       author = {{Hirschmann}, Michaela and {Dolag}, Klaus and {Saro}, Alexandro and {Bachmann}, Lisa and {Borgani}, Stefano and {Burkert}, Andreas},
        title = "{Cosmological simulations of black hole growth: AGN luminosities and downsizing}",
      journal = {\mnras},
         year = 2014,
        month = aug,
       volume = {442},
       number = {3},
        pages = {2304-2324},
          doi = {10.1093/mnras/stu1023},
archivePrefix = {arXiv},
       eprint = {1308.0333},
 primaryClass = {astro-ph.CO},
       adsurl = {https://ui.adsabs.harvard.edu/abs/2014MNRAS.442.2304H}
}

@ARTICLE{SpringelHernquist2003,
       author = {{Springel}, Volker and {Hernquist}, Lars},
        title = "{Cosmological smoothed particle hydrodynamics simulations: a hybrid multiphase model for star formation}",
      journal = {\mnras},
         year = 2003,
        month = feb,
       volume = {339},
       number = {2},
        pages = {289-311},
          doi = {10.1046/j.1365-8711.2003.06206.x},
archivePrefix = {arXiv},
       eprint = {astro-ph/0206393},
 primaryClass = {astro-ph},
       adsurl = {https://ui.adsabs.harvard.edu/abs/2003MNRAS.339..289S}
}

@ARTICLE{2014arXiv1412.6533A,
       author = {{Arth}, Alexander and {Dolag}, Klaus and {Beck}, Alexander M. and {Petkova}, Margarita and {Lesch}, Harald},
        title = "{Anisotropic thermal conduction in galaxy clusters with MHD in Gadget}",
      journal = {arXiv e-prints},
         year = 2014,
        month = dec,
          eid = {arXiv:1412.6533},
        pages = {arXiv:1412.6533},
          doi = {10.48550/arXiv.1412.6533},
archivePrefix = {arXiv},
       eprint = {1412.6533},
 primaryClass = {astro-ph.CO},
       adsurl = {https://ui.adsabs.harvard.edu/abs/2014arXiv1412.6533A}
}

@ARTICLE{Bardeen1972,
       author = {{Bardeen}, James M. and {Press}, William H. and {Teukolsky}, Saul A.},
        title = "{Rotating Black Holes: Locally Nonrotating Frames, Energy Extraction, and Scalar Synchrotron Radiation}",
      journal = {\apj},
         year = 1972,
        month = dec,
       volume = {178},
        pages = {347-370},
          doi = {10.1086/151796},
       adsurl = {https://ui.adsabs.harvard.edu/abs/1972ApJ...178..347B}
}

@ARTICLE{Aird2012,
       author = {{Aird}, James and {Coil}, Alison L. and {Moustakas}, John and {Blanton}, Michael R. and {Burles}, Scott M. and {Cool}, Richard J. and {Eisenstein}, Daniel J. and {Smith}, M. Stephen M. and {Wong}, Kenneth C. and {Zhu}, Guangtun},
        title = "{PRIMUS: The Dependence of AGN Accretion on Host Stellar Mass and Color}",
      journal = {\apj},
         year = 2012,
        month = feb,
       volume = {746},
       number = {1},
          eid = {90},
        pages = {90},
          doi = {10.1088/0004-637X/746/1/90},
archivePrefix = {arXiv},
       eprint = {1107.4368},
 primaryClass = {astro-ph.CO},
       adsurl = {https://ui.adsabs.harvard.edu/abs/2012ApJ...746...90A}
}

@ARTICLE{Angus2026,
       author = {{Angus}, C.~R. and {Smith}, A.~J. and {Magill}, D. and {Ramsden}, P. and {Sarin}, N. and {Nicholl}, M. and {Mockler}, B. and {Hammerstein}, E. and {Stein}, R. and {Yao}, Y. and {de Boer}, T. and {Chambers}, K.~C. and {Huber}, M.~E. and {Lin}, C.-C. and {Lowe}, T.~B. and {Magnier}, E.~A. and {Smartt}, S.~J. and {Wainscoat}, R.~J.},
        title = "{Can tidal disruption event models reliably measure black hole masses?}",
      journal = {\mnras},
         year = 2026,
        month = jul,
          doi = {10.1093/mnras/stag1285},
archivePrefix = {arXiv},
       eprint = {2601.04406},
 primaryClass = {astro-ph.HE},
       adsurl = {https://ui.adsabs.harvard.edu/abs/2026MNRAS.tmp.1210A}
}

@ARTICLE{Begelman1980,
       author = {{Begelman}, M.~C. and {Blandford}, R.~D. and {Rees}, M.~J.},
        title = "{Massive black hole binaries in active galactic nuclei}",
      journal = {\nat},
         year = 1980,
        month = sep,
       volume = {287},
       number = {5780},
        pages = {307-309},
          doi = {10.1038/287307a0},
       adsurl = {https://ui.adsabs.harvard.edu/abs/1980Natur.287..307B}
}

@ARTICLE{Boyer1967,
       author = {{Boyer}, Robert H. and {Lindquist}, Richard W.},
        title = "{Maximal Analytic Extension of the Kerr Metric}",
      journal = {Journal of Mathematical Physics},
         year = 1967,
        month = feb,
       volume = {8},
       number = {2},
        pages = {265-281},
          doi = {10.1063/1.1705193},
       adsurl = {https://ui.adsabs.harvard.edu/abs/1967JMP.....8..265B}
}

@ARTICLE{Bricman2020,
       author = {{Bricman}, Katja and {Gomboc}, Andreja},
        title = "{The Prospects of Observing Tidal Disruption Events with the Large Synoptic Survey Telescope}",
      journal = {\apj},
         year = 2020,
        month = feb,
       volume = {890},
       number = {1},
          eid = {73},
        pages = {73},
          doi = {10.3847/1538-4357/ab6989},
archivePrefix = {arXiv},
       eprint = {1906.08235},
 primaryClass = {astro-ph.HE},
       adsurl = {https://ui.adsabs.harvard.edu/abs/2020ApJ...890...73B}
}

@software{Carlsson2022,
  author    = {{Carlsson}, Kristoffer and {Karrasch}, Daniel and {Bauer}, Nicholas and {Samuel}, Okon and {Kelman}, Tony and {Schmerling}, Ed and {Hoffimann}, J{\'u}lio and {Visser}, Martijn and {San-Jose}, Pablo and {Christie}, Josh and {Ferris}, Andy and {Anthony Blaom}, PhD and {Pasquier}, Benoit and {Lucibello}, Carlo and {C42f} and {Carvalho}, Elias and {Saba}, Elliot and {Goretkin}, Gustavo and {Orson}, Ilya and {Choudhury}, Shushman and {Nagy}, Tamas},
  title     = {{KristofferC/NearestNeighbors.jl: v0.4.13}},
  year      = 2022,
  month     = dec,
  eid       = {10.5281/zenodo.7468949},
  doi       = {10.5281/zenodo.7468949},
  version   = {v0.4.13},
  publisher = {Zenodo},
  adsurl    = {https://ui.adsabs.harvard.edu/abs/2022zndo...7468949C}
}

@ARTICLE{Chowdhury2026,
       author = {{Chowdhury}, Rudrani Kar and {Dai}, Lixin and {Chang}, Janet N.~Y. and {Chan}, Tsang Keung},
        title = "{TDEs on FIRE: Illuminating the Cosmic Evolution of Tidal Disruption Rates}",
      journal = {arXiv e-prints},
         year = 2026,
        month = jun,
          eid = {arXiv:2606.04740},
        pages = {arXiv:2606.04740},
          doi = {10.48550/arXiv.2606.04740},
archivePrefix = {arXiv},
       eprint = {2606.04740},
 primaryClass = {astro-ph.HE},
       adsurl = {https://ui.adsabs.harvard.edu/abs/2026arXiv260604740C}
}

@ARTICLE{Dolag2009,
       author = {{Dolag}, K. and {Borgani}, S. and {Murante}, G. and {Springel}, V.},
        title = "{Substructures in hydrodynamical cluster simulations}",
      journal = {\mnras},
         year = 2009,
        month = oct,
       volume = {399},
       number = {2},
        pages = {497-514},
          doi = {10.1111/j.1365-2966.2009.15034.x},
archivePrefix = {arXiv},
       eprint = {0808.3401},
 primaryClass = {astro-ph},
       adsurl = {https://ui.adsabs.harvard.edu/abs/2009MNRAS.399..497D}
}

@article{Dolag2023,
  author        = {{Dolag}, Klaus and {Sorce}, Jenny G. and {Pilipenko}, Sergey and {Hern{\'a}ndez-Mart{\'\i}nez}, Elena and {Valentini}, Milena and {Gottl{\"o}ber}, Stefan and {Aghanim}, Nabila and {Khabibullin}, Ildar},
  title         = {{Simulating the LOcal Web (SLOW). I. Anomalies in the local density field}},
  journal       = {\aap},
  year          = 2023,
  month         = sep,
  volume        = {677},
  eid           = {A169},
  pages         = {A169},
  doi           = {10.1051/0004-6361/202346213},
  archiveprefix = {arXiv},
  eprint        = {2302.10960},
  primaryclass  = {astro-ph.CO},
  adsurl        = {https://ui.adsabs.harvard.edu/abs/2023A&A...677A.169D}
}

@ARTICLE{Dolag2025,
       author = {{Dolag}, Klaus and {Remus}, Rhea-Silvia and {Valenzuela}, Lucas M. and {Kimmig}, Lucas C. and {Seidel}, Benjamin and {Fortune}, Silvio and {Stoiber}, Johannes and {Ivleva}, Anna and {Hoffmann}, Tadziu and {Biffi}, Veronica and {Marini}, Ilaria and {Popesso}, Paola and {Vladutescu-Zopp}, Stephan},
        title = "{Encyclopedia Magneticum: Scaling Relations from Cosmic Dawn to Present Day}",
      journal = {arXiv e-prints},
         year = 2025,
        month = apr,
          eid = {arXiv:2504.01061},
        pages = {arXiv:2504.01061},
          doi = {10.48550/arXiv.2504.01061},
archivePrefix = {arXiv},
       eprint = {2504.01061},
 primaryClass = {astro-ph.CO},
       adsurl = {https://ui.adsabs.harvard.edu/abs/2025arXiv250401061D}
}

@ARTICLE{Evans1989,
       author = {{Evans}, Charles R. and {Kochanek}, Christopher S.},
        title = "{The Tidal Disruption of a Star by a Massive Black Hole}",
      journal = {\apjl},
         year = 1989,
        month = nov,
       volume = {346},
        pages = {L13},
          doi = {10.1086/185567},
       adsurl = {https://ui.adsabs.harvard.edu/abs/1989ApJ...346L..13E}
}

@ARTICLE{Faber1997,
       author = {{Faber}, S.~M. and {Tremaine}, Scott and {Ajhar}, Edward A. and {Byun}, Yong-Ik and {Dressler}, Alan and {Gebhardt}, Karl and {Grillmair}, Carl and {Kormendy}, John and {Lauer}, Tod R. and {Richstone}, Douglas},
        title = "{The Centers of Early-Type Galaxies with HST. IV. Central Parameter Relations.}",
      journal = {\aj},
         year = 1997,
        month = nov,
       volume = {114},
        pages = {1771},
          doi = {10.1086/118606},
archivePrefix = {arXiv},
       eprint = {astro-ph/9610055},
 primaryClass = {astro-ph},
       adsurl = {https://ui.adsabs.harvard.edu/abs/1997AJ....114.1771F}
}

@ARTICLE{Ferrarese2000,
       author = {{Ferrarese}, Laura and {Merritt}, David},
        title = "{A Fundamental Relation between Supermassive Black Holes and Their Host Galaxies}",
      journal = {\apjl},
         year = 2000,
        month = aug,
       volume = {539},
       number = {1},
        pages = {L9-L12},
          doi = {10.1086/312838},
archivePrefix = {arXiv},
       eprint = {astro-ph/0006053},
 primaryClass = {astro-ph},
       adsurl = {https://ui.adsabs.harvard.edu/abs/2000ApJ...539L...9F}
}

@ARTICLE{Frank1976,
       author = {{Frank}, J. and {Rees}, M.~J.},
        title = "{Effects of massive black holes on dense stellar systems.}",
      journal = {\mnras},
         year = 1976,
        month = sep,
       volume = {176},
        pages = {633-647},
          doi = {10.1093/mnras/176.3.633},
       adsurl = {https://ui.adsabs.harvard.edu/abs/1976MNRAS.176..633F}
}

@ARTICLE{French2026,
       author = {{French}, K. Decker and {Mockler}, Brenna and {Earl}, Nicholas and {Murphey}, Tanner},
        title = "{Prospects for Measuring Black Hole Masses using TDEs with the Vera C. Rubin Observatory}",
      journal = {\pasp},
         year = 2026,
        month = jan,
       volume = {138},
       number = {1},
          eid = {014101},
        pages = {014101},
          doi = {10.1088/1538-3873/ae2cca},
archivePrefix = {arXiv},
       eprint = {2512.13409},
 primaryClass = {astro-ph.GA},
       adsurl = {https://ui.adsabs.harvard.edu/abs/2026PASP..138a4101F}
}

@ARTICLE{Gebhardt2000,
       author = {{Gebhardt}, Karl and {Bender}, Ralf and {Bower}, Gary and {Dressler}, Alan and {Faber}, S.~M. and {Filippenko}, Alexei V. and {Green}, Richard and {Grillmair}, Carl and {Ho}, Luis C. and {Kormendy}, John and {Lauer}, Tod R. and {Magorrian}, John and {Pinkney}, Jason and {Richstone}, Douglas and {Tremaine}, Scott},
        title = "{A Relationship between Nuclear Black Hole Mass and Galaxy Velocity Dispersion}",
      journal = {\apjl},
         year = 2000,
        month = aug,
       volume = {539},
       number = {1},
        pages = {L13-L16},
          doi = {10.1086/312840},
archivePrefix = {arXiv},
       eprint = {astro-ph/0006289},
 primaryClass = {astro-ph},
       adsurl = {https://ui.adsabs.harvard.edu/abs/2000ApJ...539L..13G}
}

@ARTICLE{Gorski2005,
       author = {{G{\'o}rski}, K.~M. and {Hivon}, E. and {Banday}, A.~J. and {Wandelt}, B.~D. and {Hansen}, F.~K. and {Reinecke}, M. and {Bartelmann}, M.},
        title = "{HEALPix: A Framework for High-Resolution Discretization and Fast Analysis of Data Distributed on the Sphere}",
      journal = {\apj},
         year = 2005,
        month = apr,
       volume = {622},
       number = {2},
        pages = {759-771},
          doi = {10.1086/427976},
archivePrefix = {arXiv},
       eprint = {astro-ph/0409513},
 primaryClass = {astro-ph},
       adsurl = {https://ui.adsabs.harvard.edu/abs/2005ApJ...622..759G}
}

@ARTICLE{Groth2023,
       author = {{Groth}, Frederick and {Steinwandel}, Ulrich P. and {Valentini}, Milena and {Dolag}, Klaus},
        title = "{The cosmological simulation code OPENGADGET3 - implementation of meshless finite mass}",
      journal = {\mnras},
         year = 2023,
        month = nov,
       volume = {526},
       number = {1},
        pages = {616-644},
          doi = {10.1093/mnras/stad2717},
archivePrefix = {arXiv},
       eprint = {2301.03612},
 primaryClass = {astro-ph.IM},
       adsurl = {https://ui.adsabs.harvard.edu/abs/2023MNRAS.526..616G}
}

@ARTICLE{Grotova2025,
       author = {{Grotova}, I. and {Rau}, A. and {Baldini}, P. and {Goodwin}, A.~J. and {Liu}, Z. and {Merloni}, A. and {Salvato}, M. and {Anderson}, G.~E. and {Arcodia}, R. and {Buchner}, J. and {Krumpe}, M. and {Malyali}, A. and {Masterson}, M. and {Miller-Jones}, J.~C.~A. and {Nandra}, K. and {Shirley}, R.},
        title = "{The population of tidal disruption events discovered with eROSITA}",
      journal = {\aap},
         year = 2025,
        month = may,
       volume = {697},
          eid = {A159},
        pages = {A159},
          doi = {10.1051/0004-6361/202553669},
archivePrefix = {arXiv},
       eprint = {2504.08424},
 primaryClass = {astro-ph.HE},
       adsurl = {https://ui.adsabs.harvard.edu/abs/2025A&A...697A.159G}
}

@ARTICLE{Hannah2024,
       author = {{Hannah}, Christian H. and {Seth}, Anil C. and {Stone}, Nicholas C. and {van Velzen}, Sjoert},
        title = "{Counting the Unseen. I. Nuclear Density Scaling Relations for Nucleated Galaxies}",
      journal = {\aj},
         year = 2024,
        month = sep,
       volume = {168},
       number = {3},
          eid = {137},
        pages = {137},
          doi = {10.3847/1538-3881/ad630a},
archivePrefix = {arXiv},
       eprint = {2407.10911},
 primaryClass = {astro-ph.GA},
       adsurl = {https://ui.adsabs.harvard.edu/abs/2024AJ....168..137H}
}

@ARTICLE{Hannah2025,
       author = {{Hannah}, Christian H. and {Stone}, Nicholas C. and {Seth}, Anil C. and {van Velzen}, Sjoert},
        title = "{Counting the Unseen. II. Tidal Disruption Event Rates in Nearby Galaxies with REPTiDE}",
      journal = {\apj},
         year = 2025,
        month = jul,
       volume = {988},
       number = {1},
          eid = {29},
        pages = {29},
          doi = {10.3847/1538-4357/addd1b},
archivePrefix = {arXiv},
       eprint = {2412.19935},
 primaryClass = {astro-ph.GA},
       adsurl = {https://ui.adsabs.harvard.edu/abs/2025ApJ...988...29H}
}

@ARTICLE{Hills1975,
       author = {{Hills}, J.~G.},
        title = "{Possible power source of Seyfert galaxies and QSOs}",
      journal = {\nat},
         year = 1975,
        month = mar,
       volume = {254},
       number = {5498},
        pages = {295-298},
          doi = {10.1038/254295a0},
       adsurl = {https://ui.adsabs.harvard.edu/abs/1975Natur.254..295H}
}

@ARTICLE{Holoien2016,
       author = {{Holoien}, T.~W.-S. and {Kochanek}, C.~S. and {Prieto}, J.~L. and {Stanek}, K.~Z. and {Dong}, Subo and {Shappee}, B.~J. and {Grupe}, D. and {Brown}, J.~S. and {Basu}, U. and {Beacom}, J.~F. and {Bersier}, D. and {Brimacombe}, J. and {Danilet}, A.~B. and {Falco}, E. and {Guo}, Z. and {Jose}, J. and {Herczeg}, G.~J. and {Long}, F. and {Pojmanski}, G. and {Simonian}, G.~V. and {Szczygie{\l}}, D.~M. and {Thompson}, T.~A. and {Thorstensen}, J.~R. and {Wagner}, R.~M. and {Wo{\'z}niak}, P.~R.},
        title = "{Six months of multiwavelength follow-up of the tidal disruption candidate ASASSN-14li and implied TDE rates from ASAS-SN}",
      journal = {\mnras},
         year = 2016,
        month = jan,
       volume = {455},
       number = {3},
        pages = {2918-2935},
          doi = {10.1093/mnras/stv2486},
archivePrefix = {arXiv},
       eprint = {1507.01598},
 primaryClass = {astro-ph.HE},
       adsurl = {https://ui.adsabs.harvard.edu/abs/2016MNRAS.455.2918H}
}

@ARTICLE{Ivanov2006,
       author = {{Ivanov}, P.~B. and {Chernyakova}, M.~A.},
        title = "{Relativistic cross sections of mass stripping and tidal disruption of a star by a super-massive rotating black hole}",
      journal = {\aap},
         year = 2006,
        month = mar,
       volume = {448},
       number = {3},
        pages = {843-852},
          doi = {10.1051/0004-6361:20053409},
archivePrefix = {arXiv},
       eprint = {astro-ph/0509853},
 primaryClass = {astro-ph},
       adsurl = {https://ui.adsabs.harvard.edu/abs/2006A&A...448..843I}
}

@ARTICLE{Ivleva2024,
       author = {{Ivleva}, Anna and {Remus}, Rhea-Silvia and {Valenzuela}, Lucas M. and {Dolag}, Klaus},
        title = "{Merge and strip: Dark matter-free dwarf galaxies in clusters can be formed by galaxy mergers}",
      journal = {\aap},
         year = 2024,
        month = jul,
       volume = {687},
          eid = {A105},
        pages = {A105},
          doi = {10.1051/0004-6361/202449605},
archivePrefix = {arXiv},
       eprint = {2402.09060},
 primaryClass = {astro-ph.GA},
       adsurl = {https://ui.adsabs.harvard.edu/abs/2024A&A...687A.105I}
}

@ARTICLE{Kesden2012,
       author = {{Kesden}, Michael},
        title = "{Tidal-disruption rate of stars by spinning supermassive black holes}",
      journal = {\prd},
         year = 2012,
        month = jan,
       volume = {85},
       number = {2},
          eid = {024037},
        pages = {024037},
          doi = {10.1103/PhysRevD.85.024037},
archivePrefix = {arXiv},
       eprint = {1109.6329},
 primaryClass = {astro-ph.CO},
       adsurl = {https://ui.adsabs.harvard.edu/abs/2012PhRvD..85b4037K}
}

@ARTICLE{Kewley2006,
       author = {{Kewley}, Lisa J. and {Groves}, Brent and {Kauffmann}, Guinevere and {Heckman}, Tim},
        title = "{The host galaxies and classification of active galactic nuclei}",
      journal = {\mnras},
         year = 2006,
        month = nov,
       volume = {372},
       number = {3},
        pages = {961-976},
          doi = {10.1111/j.1365-2966.2006.10859.x},
archivePrefix = {arXiv},
       eprint = {astro-ph/0605681},
 primaryClass = {astro-ph},
       adsurl = {https://ui.adsabs.harvard.edu/abs/2006MNRAS.372..961K}
}

@ARTICLE{Khabibullin2014a,
       author = {{Khabibullin}, I. and {Sazonov}, S. and {Sunyaev}, R.},
        title = "{SRG/eROSITA prospects for the detection of stellar tidal disruption flares}",
      journal = {\mnras},
         year = 2014,
        month = jan,
       volume = {437},
       number = {1},
        pages = {327-337},
          doi = {10.1093/mnras/stt1889},
archivePrefix = {arXiv},
       eprint = {1304.3376},
 primaryClass = {astro-ph.HE},
       adsurl = {https://ui.adsabs.harvard.edu/abs/2014MNRAS.437..327K}
}

@ARTICLE{Khabibullin2014b,
       author = {{Khabibullin}, I. and {Sazonov}, S.},
        title = "{Stellar tidal disruption candidates found by cross-correlating the ROSAT Bright Source Catalogue and XMM-Newton observations}",
      journal = {\mnras},
         year = 2014,
        month = oct,
       volume = {444},
       number = {2},
        pages = {1041-1053},
          doi = {10.1093/mnras/stu1491},
archivePrefix = {arXiv},
       eprint = {1407.6284},
 primaryClass = {astro-ph.HE},
       adsurl = {https://ui.adsabs.harvard.edu/abs/2014MNRAS.444.1041K}
}

@ARTICLE{Kimmig2025,
       author = {{Kimmig}, Lucas C. and {Remus}, Rhea-Silvia and {Seidel}, Benjamin and {Valenzuela}, Lucas M. and {Dolag}, Klaus and {Burkert}, Andreas},
        title = "{Blowing Out the Candle: How to Quench Galaxies at High Redshift{\textemdash}An Ensemble of Rapid Starbursts, AGN Feedback, and Environment}",
      journal = {\apj},
         year = 2025,
        month = jan,
       volume = {979},
       number = {1},
          eid = {15},
        pages = {15},
          doi = {10.3847/1538-4357/ad9472},
archivePrefix = {arXiv},
       eprint = {2310.16085},
 primaryClass = {astro-ph.GA},
       adsurl = {https://ui.adsabs.harvard.edu/abs/2025ApJ...979...15K}
}

@BOOK{Kippenhahn2013,
       author = {{Kippenhahn}, Rudolf and {Weigert}, Alfred and {Weiss}, Achim},
        title = "{Stellar Structure and Evolution}",
         year = 2013,
          doi = {10.1007/978-3-642-30304-3},
       adsurl = {https://ui.adsabs.harvard.edu/abs/2013sse..book.....K}
}

@ARTICLE{Kormendy2009,
       author = {{Kormendy}, John and {Fisher}, David B. and {Cornell}, Mark E. and {Bender}, Ralf},
        title = "{Structure and Formation of Elliptical and Spheroidal Galaxies}",
      journal = {\apjs},
         year = 2009,
        month = may,
       volume = {182},
       number = {1},
        pages = {216-309},
          doi = {10.1088/0067-0049/182/1/216},
archivePrefix = {arXiv},
       eprint = {0810.1681},
 primaryClass = {astro-ph},
       adsurl = {https://ui.adsabs.harvard.edu/abs/2009ApJS..182..216K}
}

@ARTICLE{Kormendy2013,
       author = {{Kormendy}, John and {Ho}, Luis C.},
        title = "{Coevolution (Or Not) of Supermassive Black Holes and Host Galaxies}",
      journal = {\araa},
         year = 2013,
        month = aug,
       volume = {51},
       number = {1},
        pages = {511-653},
          doi = {10.1146/annurev-astro-082708-101811},
archivePrefix = {arXiv},
       eprint = {1304.7762},
 primaryClass = {astro-ph.CO},
       adsurl = {https://ui.adsabs.harvard.edu/abs/2013ARA&A..51..511K}
}

@ARTICLE{Kudritzki2021,
       author = {{Kudritzki}, Rolf-Peter and {Teklu}, Adelheid F. and {Schulze}, Felix and {Remus}, Rhea-Silvia and {Dolag}, Klaus and {Burkert}, Andreas and {Zahid}, H. Jabran},
        title = "{Galaxy Look-back Evolution Models: A Comparison with Magneticum Cosmological Simulations and Observations}",
      journal = {\apj},
         year = 2021,
        month = apr,
       volume = {910},
       number = {2},
          eid = {87},
        pages = {87},
          doi = {10.3847/1538-4357/abe40c},
archivePrefix = {arXiv},
       eprint = {2102.04135},
 primaryClass = {astro-ph.GA},
       adsurl = {https://ui.adsabs.harvard.edu/abs/2021ApJ...910...87K}
}

@ARTICLE{Lauer2007,
       author = {{Lauer}, Tod R. and {Gebhardt}, Karl and {Faber}, S.~M. and {Richstone}, Douglas and {Tremaine}, Scott and {Kormendy}, John and {Aller}, M.~C. and {Bender}, Ralf and {Dressler}, Alan and {Filippenko}, Alexei V. and {Green}, Richard and {Ho}, Luis C.},
        title = "{The Centers of Early-Type Galaxies with Hubble Space Telescope. VI. Bimodal Central Surface Brightness Profiles}",
      journal = {\apj},
         year = 2007,
        month = jul,
       volume = {664},
       number = {1},
        pages = {226-256},
          doi = {10.1086/519229},
archivePrefix = {arXiv},
       eprint = {astro-ph/0609762},
 primaryClass = {astro-ph},
       adsurl = {https://ui.adsabs.harvard.edu/abs/2007ApJ...664..226L}
}

@ARTICLE{Magorrian1999,
       author = {{Magorrian}, John and {Tremaine}, Scott},
        title = "{Rates of tidal disruption of stars by massive central black holes}",
      journal = {\mnras},
         year = 1999,
        month = oct,
       volume = {309},
       number = {2},
        pages = {447-460},
          doi = {10.1046/j.1365-8711.1999.02853.x},
archivePrefix = {arXiv},
       eprint = {astro-ph/9902032},
 primaryClass = {astro-ph},
       adsurl = {https://ui.adsabs.harvard.edu/abs/1999MNRAS.309..447M}
}

@ARTICLE{Masterson2024,
       author = {{Masterson}, Megan and {De}, Kishalay and {Panagiotou}, Christos and {Kara}, Erin and {Arcavi}, Iair and {Eilers}, Anna-Christina and {Frostig}, Danielle and {Gezari}, Suvi and {Grotova}, Iuliia and {Liu}, Zhu and {Malyali}, Adam and {Meisner}, Aaron M. and {Merloni}, Andrea and {Newsome}, Megan and {Rau}, Arne and {Simcoe}, Robert A. and {van Velzen}, Sjoert},
        title = "{A New Population of Mid-infrared-selected Tidal Disruption Events: Implications for Tidal Disruption Event Rates and Host Galaxy Properties}",
      journal = {\apj},
         year = 2024,
        month = feb,
       volume = {961},
       number = {2},
          eid = {211},
        pages = {211},
          doi = {10.3847/1538-4357/ad18bb},
archivePrefix = {arXiv},
       eprint = {2401.01403},
 primaryClass = {astro-ph.HE},
       adsurl = {https://ui.adsabs.harvard.edu/abs/2024ApJ...961..211M}
}

@ARTICLE{McConnell2013,
       author = {{McConnell}, Nicholas J. and {Ma}, Chung-Pei},
        title = "{Revisiting the Scaling Relations of Black Hole Masses and Host Galaxy Properties}",
      journal = {\apj},
         year = 2013,
        month = feb,
       volume = {764},
       number = {2},
          eid = {184},
        pages = {184},
          doi = {10.1088/0004-637X/764/2/184},
archivePrefix = {arXiv},
       eprint = {1211.2816},
 primaryClass = {astro-ph.CO},
       adsurl = {https://ui.adsabs.harvard.edu/abs/2013ApJ...764..184M}
}

@ARTICLE{Milosavljevic2001,
       author = {{Milosavljevi{\'c}}, Milo{\v{s}} and {Merritt}, David},
        title = "{Formation of Galactic Nuclei}",
      journal = {\apj},
         year = 2001,
        month = dec,
       volume = {563},
       number = {1},
        pages = {34-62},
          doi = {10.1086/323830},
archivePrefix = {arXiv},
       eprint = {astro-ph/0103350},
 primaryClass = {astro-ph},
       adsurl = {https://ui.adsabs.harvard.edu/abs/2001ApJ...563...34M}
}

@ARTICLE{Mockler2019,
       author = {{Mockler}, Brenna and {Guillochon}, James and {Ramirez-Ruiz}, Enrico},
        title = "{Weighing Black Holes Using Tidal Disruption Events}",
      journal = {\apj},
         year = 2019,
        month = feb,
       volume = {872},
       number = {2},
          eid = {151},
        pages = {151},
          doi = {10.3847/1538-4357/ab010f},
archivePrefix = {arXiv},
       eprint = {1801.08221},
 primaryClass = {astro-ph.HE},
       adsurl = {https://ui.adsabs.harvard.edu/abs/2019ApJ...872..151M}
}

@ARTICLE{Monteiro-Oliveira2022,
       author = {{Monteiro-Oliveira}, R. and {Morell}, D.~F. and {Sampaio}, V.~M. and {Ribeiro}, A.~L.~B. and {de Carvalho}, R.~R.},
        title = "{Unveiling the internal structure of the Hercules supercluster}",
      journal = {\mnras},
         year = 2022,
        month = jan,
       volume = {509},
       number = {3},
        pages = {3470-3487},
          doi = {10.1093/mnras/stab3225},
archivePrefix = {arXiv},
       eprint = {2111.03053},
 primaryClass = {astro-ph.CO},
       adsurl = {https://ui.adsabs.harvard.edu/abs/2022MNRAS.509.3470M}
}

@ARTICLE{Mummery2024,
       author = {{Mummery}, Andrew and {van Velzen}, Sjoert and {Nathan}, Edward and {Ingram}, Adam and {Hammerstein}, Erica and {Fraser-Taliente}, Ludovic and {Balbus}, Steven},
        title = "{Fundamental scaling relationships revealed in the optical light curves of tidal disruption events}",
      journal = {\mnras},
         year = 2024,
        month = jan,
       volume = {527},
       number = {2},
        pages = {2452-2489},
          doi = {10.1093/mnras/stad3001},
archivePrefix = {arXiv},
       eprint = {2308.08255},
 primaryClass = {astro-ph.HE},
       adsurl = {https://ui.adsabs.harvard.edu/abs/2024MNRAS.527.2452M}
}

@ARTICLE{Nair2026,
       author = {{Nair}, Prajna and {Panagiotou}, Christos and {Masterson}, Megan and {De}, Kishalay and {Kara}, Erin and {Winkler}, Eleanor and {Subhi Abo Rdan}, M.},
        title = "{A Suppressed Volumetric Rate of High-Luminosity Mid-Infrared Selected Tidal Disruption Events}",
      journal = {arXiv e-prints},
         year = 2026,
        month = jun,
          eid = {arXiv:2606.31926},
        pages = {arXiv:2606.31926},
archivePrefix = {arXiv},
       eprint = {2606.31926},
 primaryClass = {astro-ph.HE},
       adsurl = {https://ui.adsabs.harvard.edu/abs/2026arXiv260631926N}
}

@ARTICLE{Patra2026,
       author = {{Patra}, Kishore C. and {Liepold}, Emily R. and {Earl}, Nicholas and {Foley}, Ryan J. and {Ma}, Chung-Pei and {Gomez}, Sebastian and {Davis}, Kyle W. and {Ramirez-Ruiz}, Enrico and {French}, K. Decker and {Walsh}, Jonelle L. and {Kaur}, Ravjit and {Taggart}, Kirsty and {Candanoza}, Joshua and {Villar}, V. Ashley and {Arunachalam}, Prasiddha and {Macias}, Phillip and {Tinyanont}, Samaporn},
        title = "{JWST and Keck observations of the off-nuclear tidal disruption event TDE 2025abcr: An evolving reprocessing layer}",
      journal = {arXiv e-prints},
         year = 2026,
        month = apr,
          eid = {arXiv:2604.16093},
        pages = {arXiv:2604.16093},
archivePrefix = {arXiv},
       eprint = {2604.16093},
 primaryClass = {astro-ph.HE},
       adsurl = {https://ui.adsabs.harvard.edu/abs/2026arXiv260416093P}
}

@article{Planck2014,
  author        = {{Planck Collaboration} and {Ade}, P.~A.~R. and {Aghanim}, N. and {Armitage-Caplan}, C. and {Arnaud}, M. and {Ashdown}, M. and {Atrio-Barandela}, F. and {Aumont}, J. and {Baccigalupi}, C. and {Banday}, A.~J. and {Barreiro}, R.~B. and {Bartlett}, J.~G. and {Battaner}, E. and {Benabed}, K. and {Beno{\^\i}t}, A. and {Benoit-L{\'e}vy}, A. and {Bernard}, J. -P. and {Bersanelli}, M. and {Bielewicz}, P. and {Bobin}, J. and {Bock}, J.~J. and {Bonaldi}, A. and {Bond}, J.~R. and {Borrill}, J. and {Bouchet}, F.~R. and {Bridges}, M. and {Bucher}, M. and {Burigana}, C. and {Butler}, R.~C. and {Calabrese}, E. and {Cappellini}, B. and {Cardoso}, J. -F. and {Catalano}, A. and {Challinor}, A. and {Chamballu}, A. and {Chary}, R. -R. and {Chen}, X. and {Chiang}, H.~C. and {Chiang}, L. -Y. and {Christensen}, P.~R. and {Church}, S. and {Clements}, D.~L. and {Colombi}, S. and {Colombo}, L.~P.~L. and {Couchot}, F. and {Coulais}, A. and {Crill}, B.~P. and {Curto}, A. and {Cuttaia}, F. and {Danese}, L. and {Davies}, R.~D. and {Davis}, R.~J. and {de Bernardis}, P. and {de Rosa}, A. and {de Zotti}, G. and {Delabrouille}, J. and {Delouis}, J. -M. and {D{\'e}sert}, F. -X. and {Dickinson}, C. and {Diego}, J.~M. and {Dolag}, K. and {Dole}, H. and {Donzelli}, S. and {Dor{\'e}}, O. and {Douspis}, M. and {Dunkley}, J. and {Dupac}, X. and {Efstathiou}, G. and {Elsner}, F. and {En{\ss}lin}, T.~A. and {Eriksen}, H.~K. and {Finelli}, F. and {Forni}, O. and {Frailis}, M. and {Fraisse}, A.~A. and {Franceschi}, E. and {Gaier}, T.~C. and {Galeotta}, S. and {Galli}, S. and {Ganga}, K. and {Giard}, M. and {Giardino}, G. and {Giraud-H{\'e}raud}, Y. and {Gjerl{\o}w}, E. and {Gonz{\'a}lez-Nuevo}, J. and {G{\'o}rski}, K.~M. and {Gratton}, S. and {Gregorio}, A. and {Gruppuso}, A. and {Gudmundsson}, J.~E. and {Haissinski}, J. and {Hamann}, J. and {Hansen}, F.~K. and {Hanson}, D. and {Harrison}, D. and {Henrot-Versill{\'e}}, S. and {Hern{\'a}ndez-Monteagudo}, C. and {Herranz}, D. and {Hildebrandt}, S.~R. and {Hivon}, E. and {Hobson}, M. and {Holmes}, W.~A. and {Hornstrup}, A. and {Hou}, Z. and {Hovest}, W. and {Huffenberger}, K.~M. and {Jaffe}, A.~H. and {Jaffe}, T.~R. and {Jewell}, J. and {Jones}, W.~C. and {Juvela}, M. and {Keih{\"a}nen}, E. and {Keskitalo}, R. and {Kisner}, T.~S. and {Kneissl}, R. and {Knoche}, J. and {Knox}, L. and {Kunz}, M. and {Kurki-Suonio}, H. and {Lagache}, G. and {L{\"a}hteenm{\"a}ki}, A. and {Lamarre}, J. -M. and {Lasenby}, A. and {Lattanzi}, M. and {Laureijs}, R.~J. and {Lawrence}, C.~R. and {Leach}, S. and {Leahy}, J.~P. and {Leonardi}, R. and {Le{\'o}n-Tavares}, J. and {Lesgourgues}, J. and {Lewis}, A. and {Liguori}, M. and {Lilje}, P.~B. and {Linden-V{\o}rnle}, M. and {L{\'o}pez-Caniego}, M. and {Lubin}, P.~M. and {Mac{\'\i}as-P{\'e}rez}, J.~F. and {Maffei}, B. and {Maino}, D. and {Mandolesi}, N. and {Maris}, M. and {Marshall}, D.~J. and {Martin}, P.~G. and {Mart{\'\i}nez-Gonz{\'a}lez}, E. and {Masi}, S. and {Massardi}, M. and {Matarrese}, S. and {Matthai}, F. and {Mazzotta}, P. and {Meinhold}, P.~R. and {Melchiorri}, A. and {Melin}, J. -B. and {Mendes}, L. and {Menegoni}, E. and {Mennella}, A. and {Migliaccio}, M. and {Millea}, M. and {Mitra}, S. and {Miville-Desch{\^e}nes}, M. -A. and {Moneti}, A. and {Montier}, L. and {Morgante}, G. and {Mortlock}, D. and {Moss}, A. and {Munshi}, D. and {Murphy}, J.~A. and {Naselsky}, P. and {Nati}, F. and {Natoli}, P. and {Netterfield}, C.~B. and {N{\o}rgaard-Nielsen}, H.~U. and {Noviello}, F. and {Novikov}, D. and {Novikov}, I. and {O'Dwyer}, I.~J. and {Osborne}, S. and {Oxborrow}, C.~A. and {Paci}, F. and {Pagano}, L. and {Pajot}, F. and {Paladini}, R. and {Paoletti}, D. and {Partridge}, B. and {Pasian}, F. and {Patanchon}, G. and {Pearson}, D. and {Pearson}, T.~J. and {Peiris}, H.~V. and {Perdereau}, O. and {Perotto}, L. and {Perrotta}, F. and {Pettorino}, V. and {Piacentini}, F. and {Piat}, M. and {Pierpaoli}, E. and {Pietrobon}, D. and {Plaszczynski}, S. and {Platania}, P. and {Pointecouteau}, E. and {Polenta}, G. and {Ponthieu}, N. and {Popa}, L. and {Poutanen}, T. and {Pratt}, G.~W. and {Pr{\'e}zeau}, G. and {Prunet}, S. and {Puget}, J. -L. and {Rachen}, J.~P. and {Reach}, W.~T. and {Rebolo}, R. and {Reinecke}, M. and {Remazeilles}, M. and {Renault}, C. and {Ricciardi}, S. and {Riller}, T. and {Ristorcelli}, I. and {Rocha}, G. and {Rosset}, C. and {Roudier}, G. and {Rowan-Robinson}, M. and {Rubi{\~n}o-Mart{\'\i}n}, J.~A. and {Rusholme}, B. and {Sandri}, M. and {Santos}, D. and {Savelainen}, M. and {Savini}, G. and {Scott}, D. and {Seiffert}, M.~D. and {Shellard}, E.~P.~S. and {Spencer}, L.~D. and {Starck}, J. -L. and {Stolyarov}, V. and {Stompor}, R. and {Sudiwala}, R. and {Sunyaev}, R. and {Sureau}, F. and {Sutton}, D. and {Suur-Uski}, A. -S. and {Sygnet}, J. -F. and {Tauber}, J.~A. and {Tavagnacco}, D. and {Terenzi}, L. and {Toffolatti}, L. and {Tomasi}, M. and {Tristram}, M. and {Tucci}, M. and {Tuovinen}, J. and {T{\"u}rler}, M. and {Umana}, G. and {Valenziano}, L. and {Valiviita}, J. and {Van Tent}, B. and {Vielva}, P. and {Villa}, F. and {Vittorio}, N. and {Wade}, L.~A. and {Wandelt}, B.~D. and {Wehus}, I.~K. and {White}, M. and {White}, S.~D.~M. and {Wilkinson}, A. and {Yvon}, D. and {Zacchei}, A. and {Zonca}, A.},
  title         = {{Planck 2013 results. XVI. Cosmological parameters}},
  journal       = {\aap},
  year          = 2014,
  month         = nov,
  volume        = {571},
  eid           = {A16},
  pages         = {A16},
  doi           = {10.1051/0004-6361/201321591},
  archiveprefix = {arXiv},
  eprint        = {1303.5076},
  primaryclass  = {astro-ph.CO},
  adsurl        = {https://ui.adsabs.harvard.edu/abs/2014A&A...571A..16P}
}

@ARTICLE{Polkas2024,
       author = {{Polkas}, M. and {Bonoli}, S. and {Bortolas}, E. and {Izquierdo-Villalba}, D. and {Sesana}, A. and {Broggi}, L. and {Hoyer}, N. and {Spinoso}, D.},
        title = "{Demographics of tidal disruption events with L-Galaxies: I. Volumetric TDE rates and the abundance of nuclear star clusters}",
      journal = {\aap},
         year = 2024,
        month = sep,
       volume = {689},
          eid = {A204},
        pages = {A204},
          doi = {10.1051/0004-6361/202449470},
archivePrefix = {arXiv},
       eprint = {2312.13242},
 primaryClass = {astro-ph.HE},
       adsurl = {https://ui.adsabs.harvard.edu/abs/2024A&A...689A.204P}
}

@ARTICLE{Rees1988,
       author = {{Rees}, Martin J.},
        title = "{Tidal disruption of stars by black holes of {}10$^{6}$-{}10$^{8}$ solar masses in nearby galaxies}",
      journal = {\nat},
         year = 1988,
        month = jun,
       volume = {333},
       number = {6173},
        pages = {523-528},
          doi = {10.1038/333523a0},
       adsurl = {https://ui.adsabs.harvard.edu/abs/1988Natur.333..523R}
}

@ARTICLE{Reines2015,
       author = {{Reines}, Amy E. and {Volonteri}, Marta},
        title = "{Relations between Central Black Hole Mass and Total Galaxy Stellar Mass in the Local Universe}",
      journal = {\apj},
         year = 2015,
        month = nov,
       volume = {813},
       number = {2},
          eid = {82},
        pages = {82},
          doi = {10.1088/0004-637X/813/2/82},
archivePrefix = {arXiv},
       eprint = {1508.06274},
 primaryClass = {astro-ph.GA},
       adsurl = {https://ui.adsabs.harvard.edu/abs/2015ApJ...813...82R}
}

@ARTICLE{Ryu2020,
       author = {{Ryu}, Taeho and {Krolik}, Julian and {Piran}, Tsvi},
        title = "{Measuring Stellar and Black Hole Masses of Tidal Disruption Events}",
      journal = {\apj},
         year = 2020,
        month = nov,
       volume = {904},
       number = {1},
          eid = {73},
        pages = {73},
          doi = {10.3847/1538-4357/abbf4d},
archivePrefix = {arXiv},
       eprint = {2007.13765},
 primaryClass = {astro-ph.HE},
       adsurl = {https://ui.adsabs.harvard.edu/abs/2020ApJ...904...73R}
}

@ARTICLE{Sala2024,
       author = {{Sala}, Luca and {Valentini}, Milena and {Biffi}, Veronica and {Dolag}, Klaus},
        title = "{Supermassive black hole spin evolution in cosmological simulations with OPENGADGET3}",
      journal = {\aap},
         year = 2024,
        month = may,
       volume = {685},
          eid = {A92},
        pages = {A92},
          doi = {10.1051/0004-6361/202348925},
archivePrefix = {arXiv},
       eprint = {2312.07657},
 primaryClass = {astro-ph.GA},
       adsurl = {https://ui.adsabs.harvard.edu/abs/2024A&A...685A..92S}
}

@ARTICLE{Sazonov2021,
       author = {{Sazonov}, S. and {Gilfanov}, M. and {Medvedev}, P. and {Yao}, Y. and {Khorunzhev}, G. and {Semena}, A. and {Sunyaev}, R. and {Burenin}, R. and {Lyapin}, A. and {Meshcheryakov}, A. and {Uskov}, G. and {Zaznobin}, I. and {Postnov}, K.~A. and {Dodin}, A.~V. and {Belinski}, A.~A. and {Cherepashchuk}, A.~M. and {Eselevich}, M. and {Dodonov}, S.~N. and {Grokhovskaya}, A.~A. and {Kotov}, S.~S. and {Bikmaev}, I.~F. and {Zhuchkov}, R. Ya and {Gumerov}, R.~I. and {van Velzen}, S. and {Kulkarni}, S.},
        title = "{First tidal disruption events discovered by SRG/eROSITA: X-ray/optical properties and X-ray luminosity function at z < 0.6}",
      journal = {\mnras},
         year = 2021,
        month = dec,
       volume = {508},
       number = {3},
        pages = {3820-3847},
          doi = {10.1093/mnras/stab2843},
archivePrefix = {arXiv},
       eprint = {2108.02449},
 primaryClass = {astro-ph.HE},
       adsurl = {https://ui.adsabs.harvard.edu/abs/2021MNRAS.508.3820S}
}

@ARTICLE{Schulze2011,
       author = {{Schulze}, Andreas and {Gebhardt}, Karl},
        title = "{Effect of a Dark Matter Halo on the Determination of Black Hole Masses}",
      journal = {\apj},
         year = 2011,
        month = mar,
       volume = {729},
       number = {1},
          eid = {21},
        pages = {21},
          doi = {10.1088/0004-637X/729/1/21},
archivePrefix = {arXiv},
       eprint = {1011.5077},
 primaryClass = {astro-ph.CO},
       adsurl = {https://ui.adsabs.harvard.edu/abs/2011ApJ...729...21S}
}

@ARTICLE{Schulze2018,
       author = {{Schulze}, Felix and {Remus}, Rhea-Silvia and {Dolag}, Klaus and {Burkert}, Andreas and {Emsellem}, Eric and {van de Ven}, Glenn},
        title = "{Kinematics of simulated galaxies - I. Connecting dynamical and morphological properties of early-type galaxies at different redshifts}",
      journal = {\mnras},
         year = 2018,
        month = nov,
       volume = {480},
       number = {4},
        pages = {4636-4658},
          doi = {10.1093/mnras/sty2090},
archivePrefix = {arXiv},
       eprint = {1802.01583},
 primaryClass = {astro-ph.GA},
       adsurl = {https://ui.adsabs.harvard.edu/abs/2018MNRAS.480.4636S}
}

@ARTICLE{Secker1996,
       author = {{Secker}, Jeff and {Harris}, William E.},
        title = "{The Early-Type Dwarf-to-Giant Ratio and Substructure in the Coma Cluster}",
      journal = {\apj},
         year = 1996,
        month = oct,
       volume = {469},
        pages = {623},
          doi = {10.1086/177810},
archivePrefix = {arXiv},
       eprint = {astro-ph/9605093},
 primaryClass = {astro-ph},
       adsurl = {https://ui.adsabs.harvard.edu/abs/1996ApJ...469..623S}
}

@ARTICLE{Seidel2025,
       author = {{Seidel}, B.~A. and {Dolag}, K. and {Remus}, R.-S. and {Sorce}, J.~G. and {Hern{\'a}ndez-Mart{\'\i}nez}, E. and {Khabibullin}, I. and {Aghanim}, N.},
        title = "{Simulating the LOcal Web (SLOW): IV. Not all that is close will merge in the end: Superclusters and their Lagrangian collapse regions}",
      journal = {\aap},
         year = 2025,
        month = oct,
       volume = {702},
          eid = {A243},
        pages = {A243},
          doi = {10.1051/0004-6361/202453421},
archivePrefix = {arXiv},
       eprint = {2412.08708},
 primaryClass = {astro-ph.CO},
       adsurl = {https://ui.adsabs.harvard.edu/abs/2025A&A...702A.243S}
}

@ARTICLE{Seidel2026,
       author = {{Seidel}, Benjamin and {Dolag}, Klaus and {Sorce}, Jenny G.},
        title = "{Cutting with precision -- Leveraging Collapse Volumes to generate the next generation of zoom-in initial conditions}",
      journal = {arXiv e-prints},
         year = 2026,
        month = jun,
          eid = {arXiv:2606.26230},
        pages = {arXiv:2606.26230},
          doi = {10.48550/arXiv.2606.26230},
archivePrefix = {arXiv},
       eprint = {2606.26230},
 primaryClass = {astro-ph.CO},
       adsurl = {https://ui.adsabs.harvard.edu/abs/2026arXiv260626230S}
}

@ARTICLE{Springel2001,
       author = {{Springel}, Volker and {White}, Simon D.~M. and {Tormen}, Giuseppe and {Kauffmann}, Guinevere},
        title = "{Populating a cluster of galaxies - I. Results at z=0}",
      journal = {\mnras},
         year = 2001,
        month = dec,
       volume = {328},
       number = {3},
        pages = {726-750},
          doi = {10.1046/j.1365-8711.2001.04912.x},
archivePrefix = {arXiv},
       eprint = {astro-ph/0012055},
 primaryClass = {astro-ph},
       adsurl = {https://ui.adsabs.harvard.edu/abs/2001MNRAS.328..726S}
}

@ARTICLE{Srivastav2026,
       author = {{Srivastav}, S. and {Smartt}, S.~J. and {Moore}, T. and {Smith}, K.~W. and {Young}, D.~R. and {Fulton}, M.~D. and {Angus}, C.~R. and {Nicholl}, M. and {Stevance}, H.~F. and {Chen}, T.-W. and {Pastorello}, A. and {Sommer}, J. and {Stoppa}, F. and {Tweddle}, J.~W. and {Anderson}, J.~P. and {Huber}, M.~E. and {Rest}, A. and {Rhodes}, L. and {Shingles}, L.~J. and {Aamer}, A. and {Clocchiatti}, A. and {Cooper}, A.~J. and {Erasmus}, N. and {Gillanders}, J.~H. and {Magill}, D. and {Pignata}, G. and {Ramsden}, P. and {Schmidt}, B.~P. and {Sheng}, X. and {Weston}, J.~G. and {Denneau}, L. and {Tonry}, J.~L.},
        title = "{ATLAS100 ─ I. A volume-limited sample of supernovae and related transients within 100 Mpc}",
      journal = {\mnras},
         year = 2026,
        month = jul,
       volume = {549},
       number = {4},
          eid = {stag1028},
        pages = {stag1028},
          doi = {10.1093/mnras/stag1028},
archivePrefix = {arXiv},
       eprint = {2603.03069},
 primaryClass = {astro-ph.HE},
       adsurl = {https://ui.adsabs.harvard.edu/abs/2026MNRAS.549g1028S}
}

@ARTICLE{Stone2016,
       author = {{Stone}, Nicholas C. and {Metzger}, Brian D.},
        title = "{Rates of stellar tidal disruption as probes of the supermassive black hole mass function}",
      journal = {\mnras},
         year = 2016,
        month = jan,
       volume = {455},
       number = {1},
        pages = {859-883},
          doi = {10.1093/mnras/stv2281},
archivePrefix = {arXiv},
       eprint = {1410.7772},
 primaryClass = {astro-ph.HE},
       adsurl = {https://ui.adsabs.harvard.edu/abs/2016MNRAS.455..859S}
}

@ARTICLE{Stone2020,
       author = {{Stone}, N.~C. and {Vasiliev}, E. and {Kesden}, M. and {Rossi}, E.~M. and {Perets}, H.~B. and {Amaro-Seoane}, P.},
        title = "{Rates of Stellar Tidal Disruption}",
      journal = {\ssr},
         year = 2020,
        month = mar,
       volume = {216},
       number = {3},
          eid = {35},
        pages = {35},
          doi = {10.1007/s11214-020-00651-4},
archivePrefix = {arXiv},
       eprint = {2003.08953},
 primaryClass = {astro-ph.HE},
       adsurl = {https://ui.adsabs.harvard.edu/abs/2020SSRv..216...35S}
}

@ARTICLE{Stoiber2025,
       author = {{Stoiber}, Johannes and {Valenzuela}, Lucas M. and {Remus}, Rhea-Silvia and {Kimmig}, Lucas C. and {Pippert}, Jan-Niklas and {Sola}, Elisabeth and {Dolag}, Klaus},
        title = "{Living the stream: Properties and progenitors of tidal shells and streams around galaxies from Magneticum}",
      journal = {arXiv e-prints},
         year = 2025,
        month = sep,
          eid = {arXiv:2509.25307},
        pages = {arXiv:2509.25307},
          doi = {10.48550/arXiv.2509.25307},
archivePrefix = {arXiv},
       eprint = {2509.25307},
 primaryClass = {astro-ph.GA},
       adsurl = {https://ui.adsabs.harvard.edu/abs/2025arXiv250925307S}
}

@ARTICLE{Teklu2015,
       author = {{Teklu}, Adelheid F. and {Remus}, Rhea-Silvia and {Dolag}, Klaus and {Beck}, Alexander M. and {Burkert}, Andreas and {Schmidt}, Andreas S. and {Schulze}, Felix and {Steinborn}, Lisa K.},
        title = "{Connecting Angular Momentum and Galactic Dynamics: The Complex Interplay between Spin, Mass, and Morphology}",
      journal = {\apj},
         year = 2015,
        month = oct,
       volume = {812},
       number = {1},
          eid = {29},
        pages = {29},
          doi = {10.1088/0004-637X/812/1/29},
archivePrefix = {arXiv},
       eprint = {1503.03501},
 primaryClass = {astro-ph.GA},
       adsurl = {https://ui.adsabs.harvard.edu/abs/2015ApJ...812...29T}
}

@ARTICLE{Teklu2017,
       author = {{Teklu}, Adelheid F. and {Remus}, Rhea-Silvia and {Dolag}, Klaus and {Burkert}, Andreas},
        title = "{The morphology-density relation: impact on the satellite fraction}",
      journal = {\mnras},
         year = 2017,
        month = dec,
       volume = {472},
       number = {4},
        pages = {4769-4785},
          doi = {10.1093/mnras/stx2303},
archivePrefix = {arXiv},
       eprint = {1702.06546},
 primaryClass = {astro-ph.GA},
       adsurl = {https://ui.adsabs.harvard.edu/abs/2017MNRAS.472.4769T}
}

@article{Tully2013,
  author        = {{Tully}, R. Brent and {Courtois}, H{\'e}l{\`e}ne M. and {Dolphin}, Andrew E. and {Fisher}, J. Richard and {H{\'e}raudeau}, Philippe and {Jacobs}, Bradley A. and {Karachentsev}, Igor D. and {Makarov}, Dmitry and {Makarova}, Lidia and {Mitronova}, Sofia and {Rizzi}, Luca and {Shaya}, Edward J. and {Sorce}, Jenny G. and {Wu}, Po-Feng},
  title         = {{Cosmicflows-2: The Data}},
  journal       = {\aj},
  year          = 2013,
  month         = oct,
  volume        = {146},
  number        = {4},
  eid           = {86},
  pages         = {86},
  doi           = {10.1088/0004-6256/146/4/86},
  archiveprefix = {arXiv},
  eprint        = {1307.7213},
  primaryclass  = {astro-ph.CO},
  adsurl        = {https://ui.adsabs.harvard.edu/abs/2013AJ....146...86T}
}

@ARTICLE{vanVelzen2014,
       author = {{van Velzen}, Sjoert and {Farrar}, Glennys R.},
        title = "{Measurement of the Rate of Stellar Tidal Disruption Flares}",
      journal = {\apj},
         year = 2014,
        month = sep,
       volume = {792},
       number = {1},
          eid = {53},
        pages = {53},
          doi = {10.1088/0004-637X/792/1/53},
archivePrefix = {arXiv},
       eprint = {1407.6425},
 primaryClass = {astro-ph.GA},
       adsurl = {https://ui.adsabs.harvard.edu/abs/2014ApJ...792...53V}
}

@ARTICLE{Wang2004,
       author = {{Wang}, Jianxiang and {Merritt}, David},
        title = "{Revised Rates of Stellar Disruption in Galactic Nuclei}",
      journal = {\apj},
         year = 2004,
        month = jan,
       volume = {600},
       number = {1},
        pages = {149-161},
          doi = {10.1086/379767},
archivePrefix = {arXiv},
       eprint = {astro-ph/0305493},
 primaryClass = {astro-ph},
       adsurl = {https://ui.adsabs.harvard.edu/abs/2004ApJ...600..149W}
}

@ARTICLE{Wang2018,
       author = {{Wang}, Tinggui and {Yan}, Lin and {Dou}, Liming and {Jiang}, Ning and {Sheng}, Zhenfeng and {Yang}, Chenwei},
        title = "{Long-term decline of the mid-infrared emission of normal galaxies: dust echo of tidal disruption flare?}",
      journal = {\mnras},
         year = 2018,
        month = jul,
       volume = {477},
       number = {3},
        pages = {2943-2965},
          doi = {10.1093/mnras/sty465},
archivePrefix = {arXiv},
       eprint = {1802.05105},
 primaryClass = {astro-ph.HE},
       adsurl = {https://ui.adsabs.harvard.edu/abs/2018MNRAS.477.2943W}
}

@ARTICLE{Yao2023,
       author = {{Yao}, Yuhan and {Ravi}, Vikram and {Gezari}, Suvi and {van Velzen}, Sjoert and {Lu}, Wenbin and {Schulze}, Steve and {Somalwar}, Jean J. and {Kulkarni}, S.~R. and {Hammerstein}, Erica and {Nicholl}, Matt and {Graham}, Matthew J. and {Perley}, Daniel A. and {Cenko}, S. Bradley and {Stein}, Robert and {Ricarte}, Angelo and {Chadayammuri}, Urmila and {Quataert}, Eliot and {Bellm}, Eric C. and {Bloom}, Joshua S. and {Dekany}, Richard and {Drake}, Andrew J. and {Groom}, Steven L. and {Mahabal}, Ashish A. and {Prince}, Thomas A. and {Riddle}, Reed and {Rusholme}, Ben and {Sharma}, Yashvi and {Sollerman}, Jesper and {Yan}, Lin},
        title = "{Tidal Disruption Event Demographics with the Zwicky Transient Facility: Volumetric Rates, Luminosity Function, and Implications for the Local Black Hole Mass Function}",
      journal = {\apjl},
         year = 2023,
        month = sep,
       volume = {955},
       number = {1},
          eid = {L6},
        pages = {L6},
          doi = {10.3847/2041-8213/acf216},
archivePrefix = {arXiv},
       eprint = {2303.06523},
 primaryClass = {astro-ph.HE},
       adsurl = {https://ui.adsabs.harvard.edu/abs/2023ApJ...955L...6Y}
}

@ARTICLE{Zoeller2025,
       author = {{Z{\"o}ller}, Raphael and {Kluge}, Matthias and {Bender}, Ralf and {Pippert}, Jan-Niklas and {Seidel}, Benjamin and {G{\"o}ssl}, Claus and {Hopp}, Ulrich and {Kellermann}, Hanna and {Ries}, Christoph and {Riffeser}, Arno and {Schmidt}, Michael and {Thomas}, Luis},
        title = "{Galaxy Luminosity Function of the Coma Cluster from Deep $u'-g'-r'$ Wendelstein Imaging Data}",
      journal = {arXiv e-prints},
         year = 2025,
        month = oct,
          eid = {arXiv:2510.26889},
        pages = {arXiv:2510.26889},
          doi = {10.48550/arXiv.2510.26889},
archivePrefix = {arXiv},
       eprint = {2510.26889},
 primaryClass = {astro-ph.GA},
       adsurl = {https://ui.adsabs.harvard.edu/abs/2025arXiv251026889Z}
}


\begin{appendix} 

\section{Supplementary figures}

\begin{figure}[h!]
    \centering
    \includegraphics[width=\linewidth]{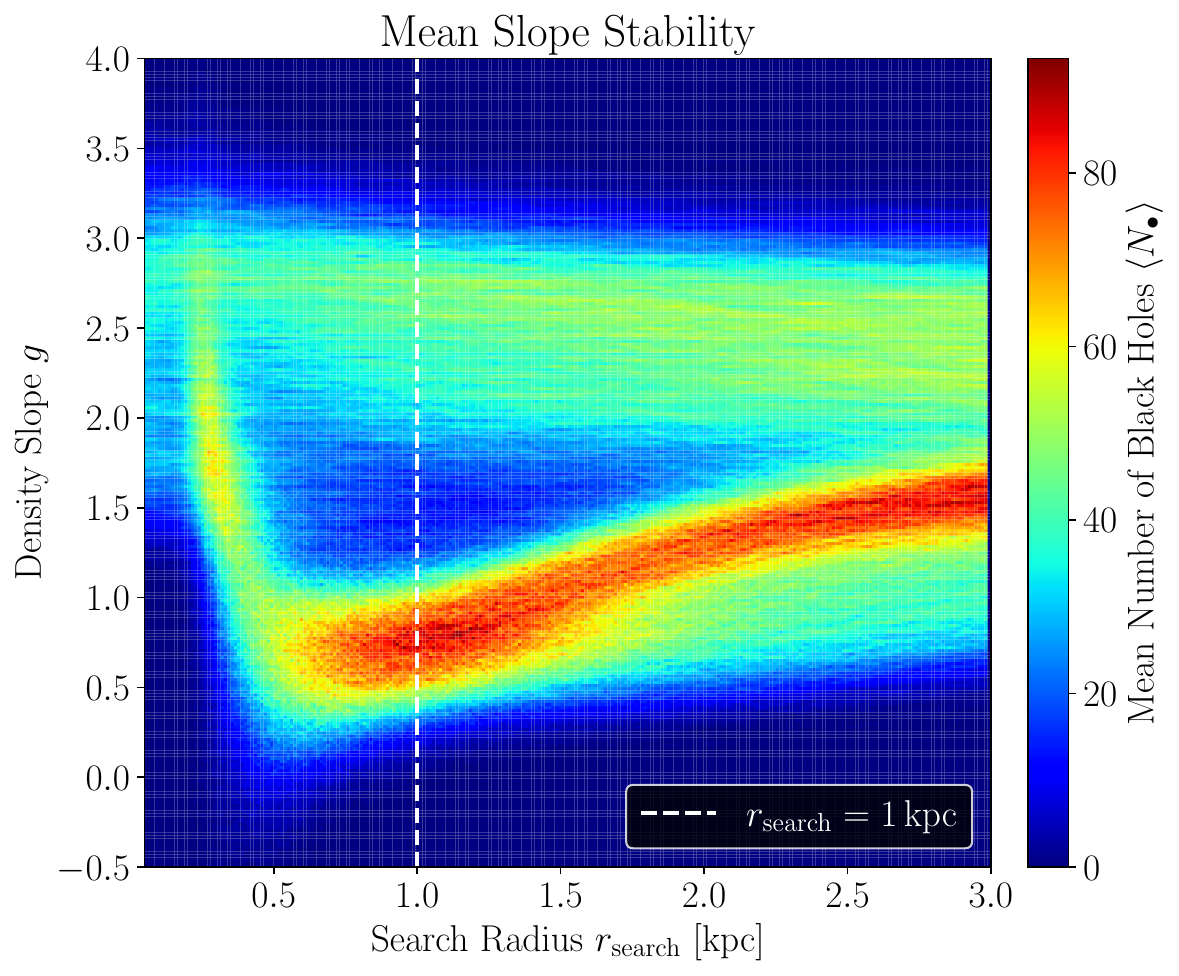}
    \caption{Stability map of the search radius $\rsearch$ denoting the radial distance in which stars are considered for the linear fit of the stellar density slope $g$. The map shows numerical instabilities at $\rsearch\lesssim 0.5\,\kpc$ and progresses into a clear bimodal distribution at $\rsearch\gtrsim 0.5\,\kpc$, which we adopt for analysis at $\rsearch = 1\,\kpc$. This bimodal distribution fades at large radii, where $\rsearch$ starts capturing stellar particles of entire galaxies.}
    \label{fig:slope_rcap_map}
\end{figure}

The following figures provide additional context for the analysis presented in the main text. Figure \ref{fig:slope_rcap_map} shows the stability map of the stellar density slope for different search radii. Depending on the stellar particles enclosed within $\rsearch$, the slope transitions from numerical instabilities to a clear bimodal distribution.

\begin{figure}
    \centering
    \includegraphics[width=\linewidth]{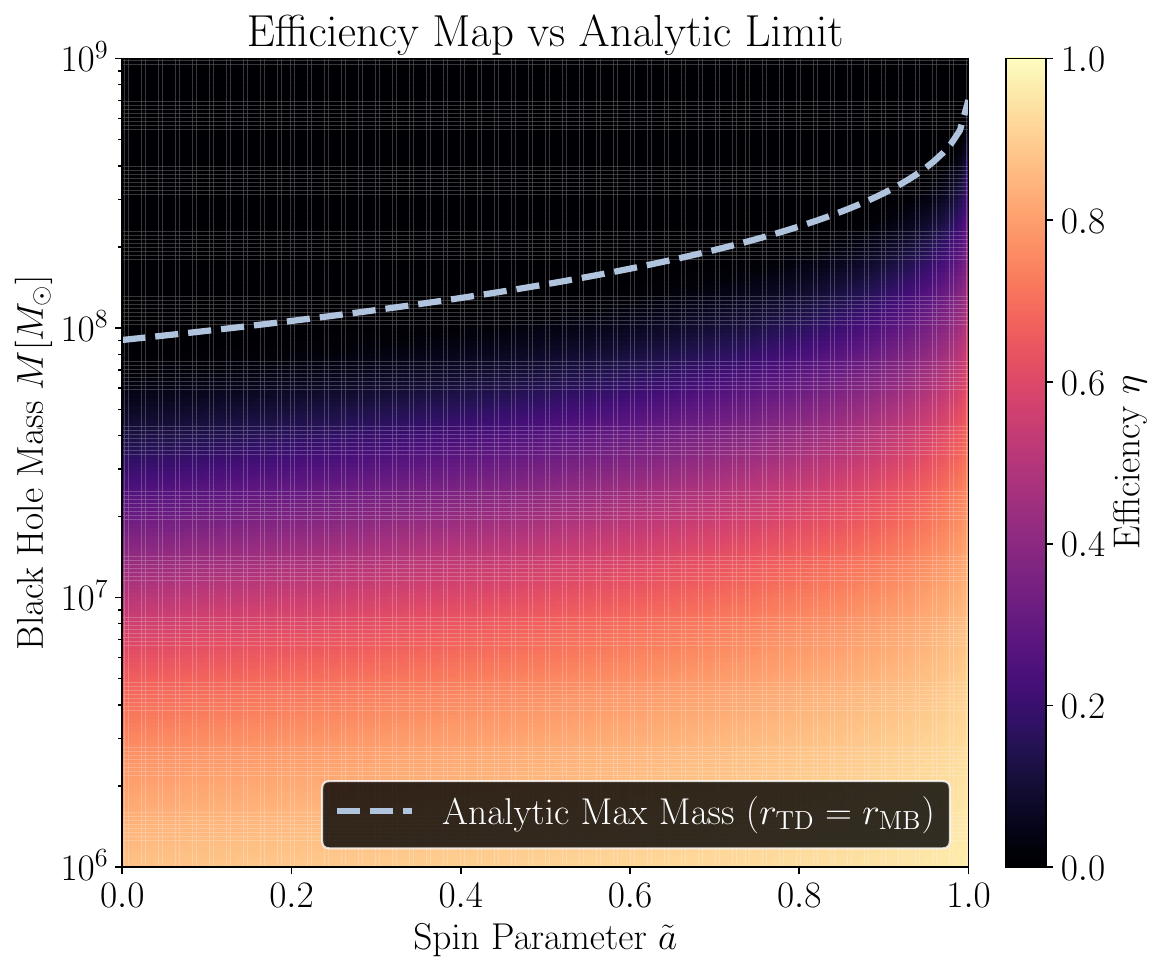}
    \caption{Relativistic tidal disruption efficiency $\eta$ evaluated across the black hole mass ($\mbh$) and the dimensionless spin ($\atilde$) parameter space. The colormap indicates the fraction of the loss cone that results in an observable TDE, avoiding direct capture by the black hole. The overlaid analytical curve represents the maximum black hole mass $M_{\bullet,\,\mathrm{max}}$ capable of producing a TDE.}
    \label{fig:kesden_efficiency_map}
\end{figure}

A heatmap of the Kesden efficiency per $\tilde{a}$-$\mbh$ bin is depicted in Figure \ref{fig:kesden_efficiency_map}. A gradual reduction of $\eta$ can be observed towards the analytical Hills mass limit across all spins, where fast spinning black holes can reach higher masses while capable of tidally disrupting stars.

\begin{figure}
    \centering
    \includegraphics[width=\linewidth]{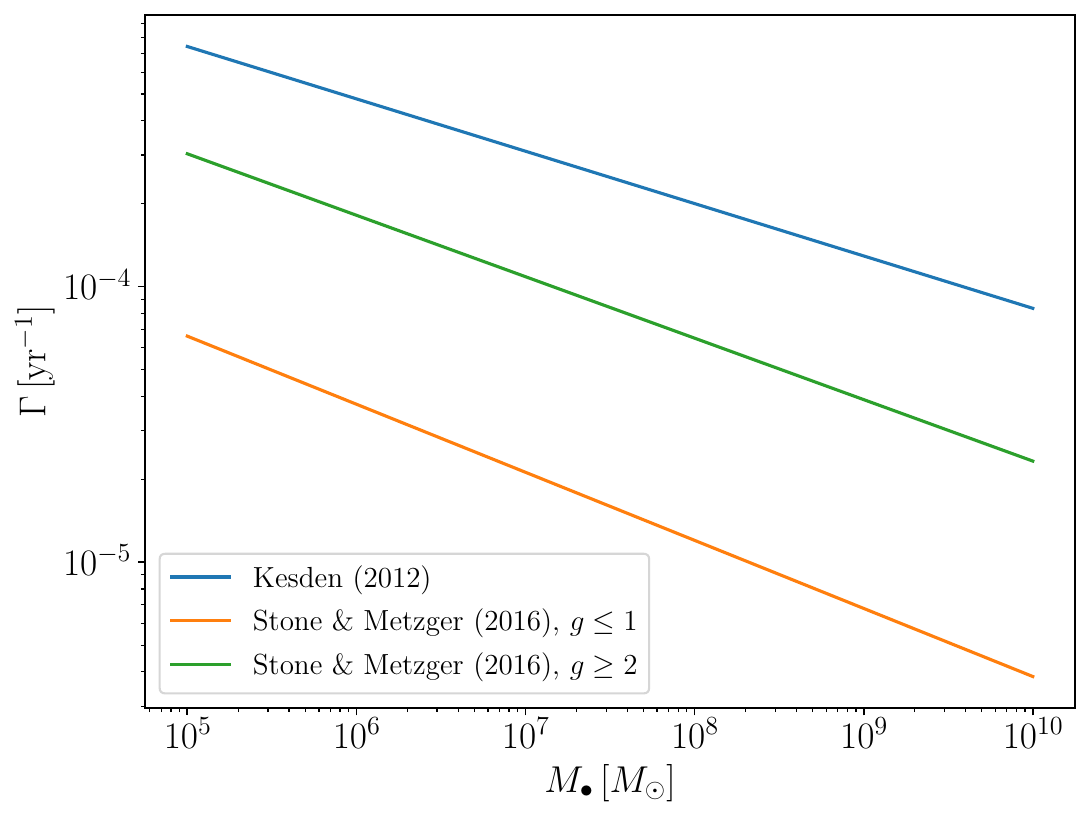}
    \caption{TDE baseline rates $\Gamma$ as a function of black hole mass $\mbh$. The SM16 baseline rates for cuspy and cored galaxies are fundamentally lower for any $\mbh$ than the Kesden baseline rates.}
    \label{fig:baserate_comparison}
\end{figure}

Figure \ref{fig:baserate_comparison} shows a comparison of the baseline rates. The baseline rate stemming from the simplified assumption of isothermal spheres for stellar density profiles presented in \cite{Kesden2012} overestimates the SM16 baseline rates. The more detailed distinction between stellar density slopes provides an empirically motivated description of expected TDE rates in different stellar environments.

\begin{figure*}
    \centering
    \includegraphics[width=1.\linewidth,trim=0cm 0.5cm 0cm 0cm]{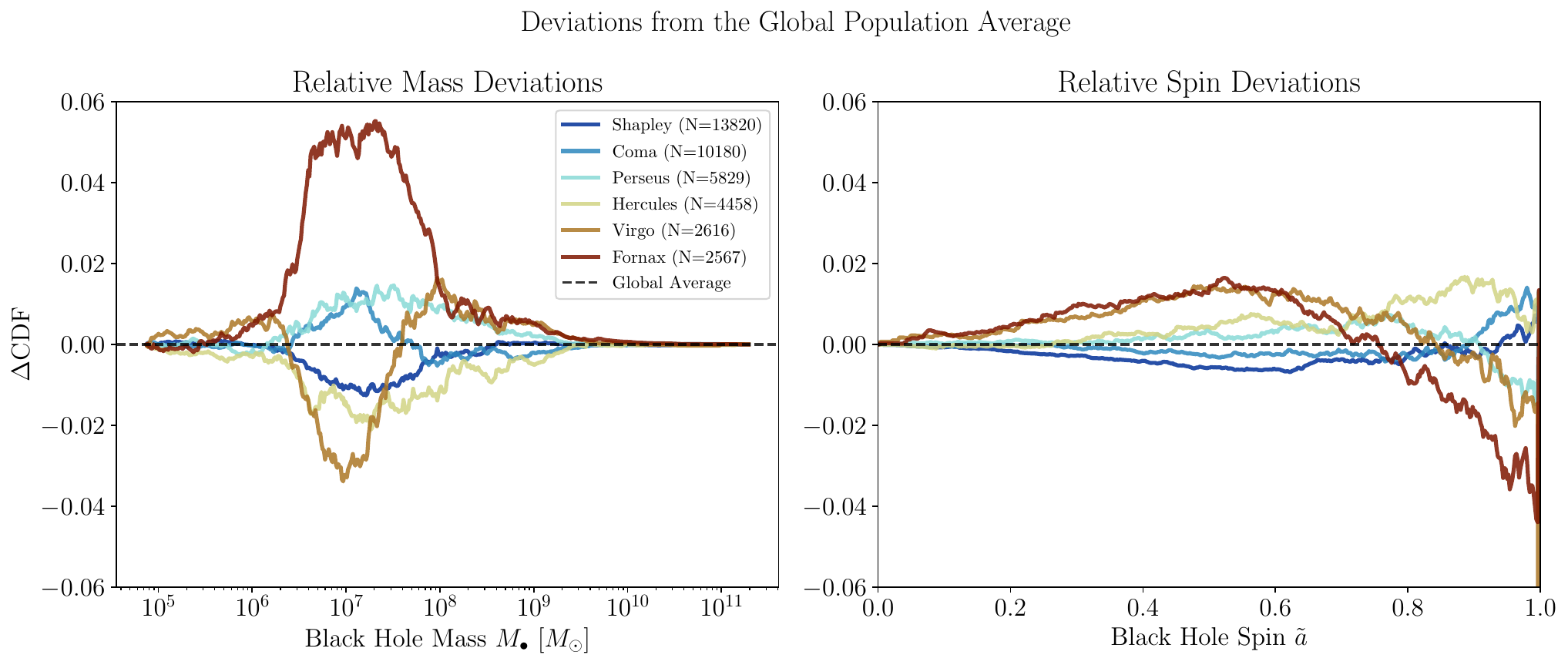}
    \caption{Relative deviations of the simulated black hole mass (left panel) and the spin (right panel) distributions from the ensemble mean. $\Delta$CDF is defined as the difference between the cumulative distribution function for an individual zoom-in volume and the ensemble average across all clusters and supercluster environments. The dashed horizontal line at $\Delta = 0$ marks the baseline at which positive or negative deviations indicate that a given cluster hosts a disproportionally higher or lower fraction of black holes compared to the overall average, respectively.}
    \label{fig:population_deviations}
\end{figure*}

Figure \ref{fig:population_deviations} displays the relative deviations of environment-specific black hole mass and spin distributions from the ensemble mean. Fornax is a clear example of hosting disproportionately more black holes below the Hills mass limit than other zoom-in regions.

\begin{figure*}
    \centering
    \includegraphics[width=\linewidth]{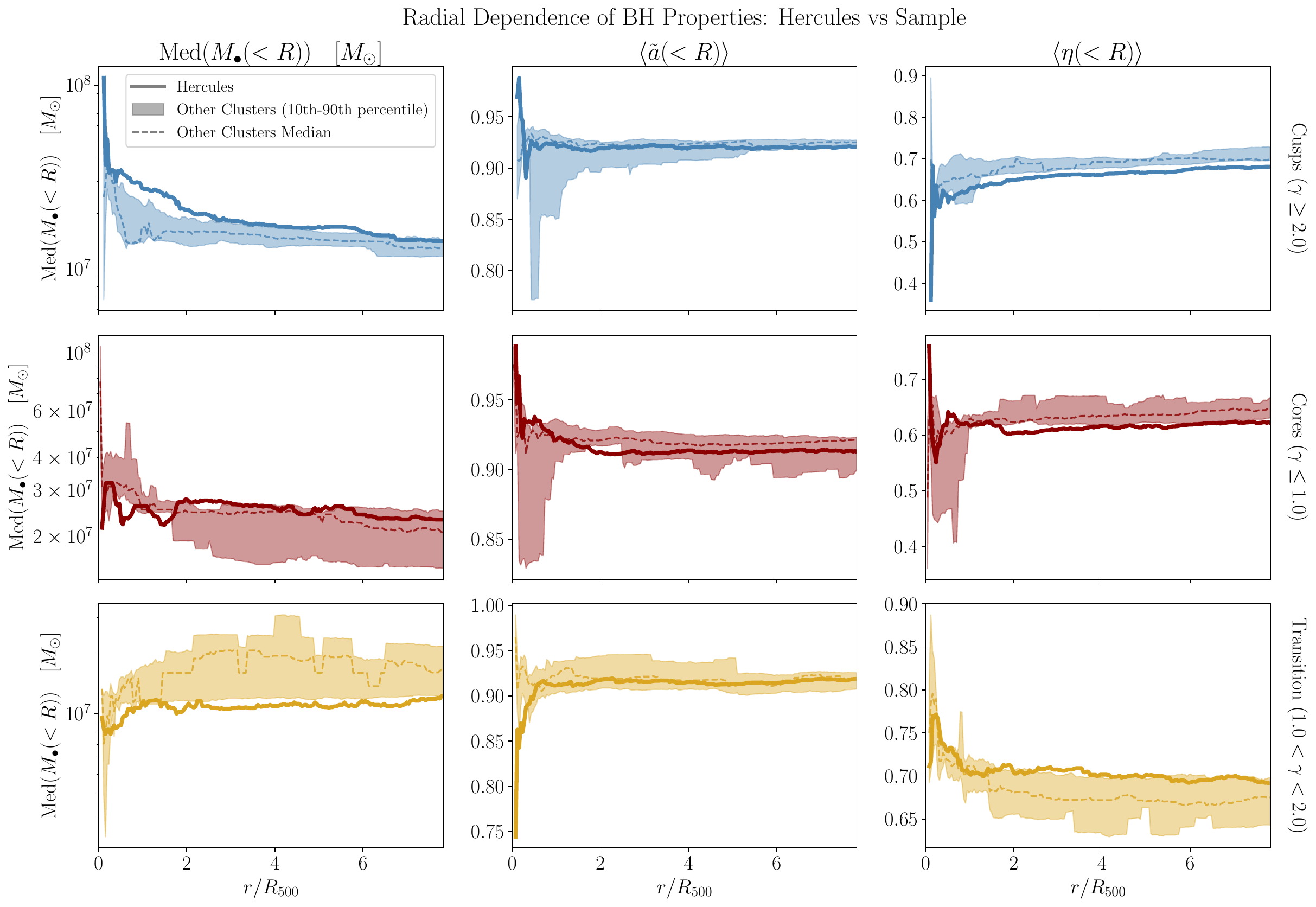}
    \caption{Cumulative radial profiles of black hole properties, comparing the Hercules supercluster to the rest of the simulated cluster sample. The rows separate host galaxies by their stellar density environments: cusps ($\gamma \ge 2.0$, top), cores ($\gamma \le 1.0$, middle), and transition environments ($1.0 < \gamma < 2.0$, bottom). The columns display the enclosed median black hole mass $\mathrm{Med}(\mbh(<R))$, the mean dimensionless spin $\langle\atilde(<R)\rangle$, and the mean Kesden efficiency $\langle\eta(<R)\rangle$ as a function of the normalized cluster radius $r/R_{500}$. In each panel, the solid line represents the Hercules supercluster, while the dashed line and shaded band indicate the median and the 10th–90th percentile range across all other simulated clusters, respectively. Hercules systematically exhibits higher enclosed black hole masses, lower dimensionless spins, and lower corresponding Kesden efficiencies, particularly within cuspy and cored environments. In transition environments, the trend appears to be the opposite.}
    \label{fig:radial_comparison_hercules}
\end{figure*}

Radial profiles of the Hercules supercluster compared to the rest of the sample are shown in Figure \ref{fig:radial_comparison_hercules}. These radial profiles provide insight into how the running median mass, spin, and Kesden efficiency evolve radially compared to the median profiles of the remaining clusters and superclusters.

\end{appendix}

\end{document}